\documentclass[showpacs,twocolumn,amsmath,superscriptaddress]{revtex4-1}
\usepackage[dvipdfmx]{graphicx}
\usepackage{color,amsmath,lmodern,bm,array,braket}

\newcommand{\ba}{{\bm a}}
\newcommand{\bk}{{\bm k}}
\newcommand{\bq}{{\bm q}}
\newcommand{\bx}{{\bm x}}
\newcommand{\by}{{\bm y}}
\newcommand{\bz}{{\bm z}}
\newcommand{\bo}{{\bm 0}}
\newcommand{\br}{{\bm r}}
\newcommand{\bmu}{\boldsymbol{\mu}}
\newcommand{\tk}{\tilde{k}}
\renewcommand{\(}{\left(}
\renewcommand{\)}{\right)}
\newcommand{\I}{{\rm I }}
\newcommand{\II}{{\rm I\hspace{-.1em}I\hspace{-.08em}}}
\newcommand{\III}{{\rm I\hspace{-.1em}I\hspace{-.1em}I }}
\newcommand{\IV}{{\rm I\hspace{-.1em}V }}

\newcommand{\be}{\begin{eqnarray}}
\newcommand{\ee}{\end{eqnarray}}
\begin{document}

\preprint{multipole superconductor}

\title{Classification of ``multipole'' superconductivity in multi-orbital systems and its implications}%

\author{T. Nomoto}%
\email{nomoto.takuya@scphys.kyoto-u.ac.jp}%
\affiliation{Department of Physics, Kyoto University, Kyoto, 606-8502, Japan}%
\author{K. Hattori}
\affiliation{Department of Physics, Tokyo Metropolitan University, Hachioji, 192-0397, Japan}%
\author{H. Ikeda}%
\affiliation{Department of Physics, Ritsumeikan University, Kusatsu, 525-8577, Japan}%

\date{\today}%

\begin{abstract}
Motivated by a growing interest in multi-orbital superconductors with spin-orbit interactions, we perform the group-theoretical classification of various unconventional superconductivity emerging in symmorphic $\rm O$, $\rm D_4$, and $\rm D_6$ space groups. The generalized Cooper pairs, which we here call ``multipole'' superconductivity, possess spin-orbital coupled (multipole) degrees of freedom, instead of the conventional spin singlet/triplet in single-orbital systems. From the classification, we obtain the following key consequences, which have been overlooked in the long history of research in this field: (1) A superconducting gap function with $\varGamma_9\otimes\varGamma_9$ in $\rm D_6$ possesses nontrivial momentum dependence, different from the usual spin 1/2 classification. (2) Unconventional gap structure can be realized in the BCS approximation of purely local (on-site) interactions irrespective of attractive/repulsive. It implies the emergence of an electron-phonon (e-ph) driven unconventional superconductivity. (3) Reflecting symmetry of orbital basis functions, there appear not symmetry-protected but inevitable line nodes/gap minima, and thus, anisotropic $s$-wave superconductivity can be naturally explained without any competitive fluctuations. 
\end{abstract}


\maketitle

\section{introduction}
In the celebrated microscopic theory by Bardeen, Cooper, and Schrieffer (BCS) in 1957~\cite{BCS}, the superconducting state is described as a condensation of Cooper pairs. The resulting Cooper pair wave function or gap function plays a role of the superconducting order parameter, which spontaneously breaks the U(1) gauge symmetry below the transition temperature $T_c$. 

The BCS theory excellently explained interesting phenomena in the traditional superconductivity. However, the class of heavy-fermion superconductors discovered around 1980~\cite{Stewart} and also the high-$T_c$ cuprates~\cite{cuprate} did not fit the BCS theory. The power-law temperature behavior in various thermodynamic quantities at low temperatures observed in these superconductors was drastically different from the conventional BCS superconductors. In the early stage, it was clear that an extension of the BCS theory is inevitable. It was soon discussed that spin-fluctuations can lead to anisotropic pairing states~\cite{Miyake,Scalapino}, in connection with superfluid $\rm ^3He$~\cite{Legget}. In such unconventional superconductivity, one or more symmetries in addition to the U(1) symmetry are broken below $T_c$. 

For instance, phase sensitive experiments such as $\pi$-junction and angle-resolved measurements clarified that the high-$T_c$ cuprates and also CeCoIn$_5$ possess the $d_{x^2-y^2}$-wave pairing state~\cite{Harlingen,ce115A,ce115B}, which belongs to $B_{1g}$ symmetry in the tetragonal crystal structure. In such case, low-energy excitations below $T_c$ are dominated by nodal quasi-particle excitations around symmetry-protected line nodes (gap zeros) on the Fermi surfaces. This situation is incompatible with the fully-gapped $s$-wave state in the conventional BCS theory. The gap structure is closely related to the pairing symmetry and the pairing mechanism. Thus, the superconducting gap function, which is one of the most fundamental quantities, continues to be hotly debated in this research field. 

In this context, group-theoretical classification of the superconducting gap functions is important and useful to investigate a variety of superconductors. Indeed, the early works~\cite{volovik1,volovik2,ueda1,sigrist1} of classification in major point groups are indispensable for the analysis of various unconventional superconductors including heavy-fermion superconductors~\cite{sigrist1,salus1,joynt1}, cuprates~\cite{annet1}, ruthenates~\cite{rice1,machida1,mackenzie1}, and so on. 

In the last decade, novel superconductors beyond these major classifications have attracted much attentions. For instance, in the non-centrosymmetric superconductors, such as CePt$_3$Si~\cite{bauer1}, UIr~\cite{akazawa1} and LaBiPt~\cite{Goll} and so on, lack of spatial-inversion (SI) symmetry admits the presence of antisymmetric spin-orbit coupling (SOC), and then the spin part of the pairing state breaks SU(2) symmetry. In this case, the so-called parity mixing occurs between spin-singlet and triplet states, which are separable under the SI symmetry and the time-reversal (TR) symmetry. Classification in these non-centrosymmetric superconductors has been established~\cite{Edelstein,gorkov1,Yip0,frigeri1,sergienko1}, and also the relation with the topological nature has been discussed~\cite{sato}. 

Regarding centrosymmetric superconductors, ``classic" heavy-fermion superconductor UPt$_3$ have attracted continuous attentions since its discovery~\cite{upt3rev1,upt3rev2}. There have been steady progress in group-theoretical considerations about the gap symmetry of $\rm UPt_3$~\cite{micklitz1,UPt3}. As Bloch states are bases of a small representation of a little group, superconducting gap functions can be also classified on the basis of the little group~\cite{izyumov1,yarzhemsky1,yarzhemsky2}. In the non-symmorphic systems, representations of a little co-group often become projective at Brillouin zone (BZ) boundary~\cite{bradley1}. This fact  yields symmetry-protected line-nodes on the BZ boundary~\cite{micklitz1}. However, it still remains unclear what type of pairing state is realized in UPt$_3$.

In the previous study based on the first-principles approach, two of the present authors found that $\rm UPt_3$ possesses the exotic multi-gap structure with twofold line-nodes, which are not allowed in the classification of a single-orbital pseudo-spin model~\cite{UPt3}. Even with only this result, we can realize the importance of classification in the multi-orbital systems. In addition, it has been gradually recognized that the multi-orbital character of gap functions is important for understanding the iron-based superconductors~\cite{lee1,hu1,ong1}. A complete set of superconducting pairing states allowed in two/three orbital models has been  summarized in Refs.\,\cite{zhou1,wan1,you1,daghofer1,fischer1}.

Thus, motivated by a growing interest in multi-orbital superconductors with spin-orbit interactions, we here perform the group-theoretical classification of various superconducting gap functions. We focus on the pairing states with zero total momentum, and demonstrate the classification of unconventional superconductivity emerging in symmorphic $\rm O$, $\rm D_4$, and $\rm D_6$ space groups. Complete sets of basis functions are summarized in several tables. Because of the SOC, multi-orbital degrees of freedom appear as multipole characters. Similarly to $\bm d$-vector in spin-triplet states, they can be specified by multipole operators in the corresponding point groups. Thus, we here call the generalized pairing state ``multipole'' superconductivity. 

From its important but complicated classification, we obtain the following key consequences, which have been overlooked in the long history of research in this field. 
\begin{enumerate}
\item A superconducting gap function with $\varGamma_9\otimes\varGamma_9$ in $\rm D_6$ possesses nontrivial momentum dependence, different from the usual spin 1/2 classification. This is related to twofold symmetric line-nodes found in the microscopic study of $\rm UPt_3$~\cite{UPt3}. 

\item Unconventional gap structure can be realized in the BCS approximation with purely local (on-site) interactions irrespective of attractive or repulsive. It implies the emergence of an electron-phonon (e-ph) driven unconventional superconductivity. Although the conventional e-ph interactions favor $s$-wave ($A_{1g}$) pairing states, the Hund's coupling and the e-ph interactions in magnetically ordered states can enhance such anisotropic pairing states. 
 
\item Reflecting the multipole characters of Cooper pairs, there appear not symmetry-protected but inevitable line nodes or gap minima, and thus, anisotropic $s$-wave superconductivity can naturally emerge without any competitive fluctuations. 
\end{enumerate}

This paper is organized as follows. In Sec.\,\ref{sec:2}, we will discuss the classification of superconducting order parameters in multi-orbital systems in terms of the local orbital bases that transform as irreducible representations of the point group in the system.
Complete tables of the Cooper pair basis functions for representative point group symmetries O, D$_4$, and D$_6$ will be demonstrated. In the final part in Sec.\,\ref{sec:2}, we will show the relations between the band-based representation and the orbital one, and clarify how the band-based Cooper pairs are related to the orbital-based ones. In Sec.\,\ref{sec:3}, we will discuss two models for the cubic $\rm O_h$ and tetragonal $\rm D_{4h}$ point groups as the applications of the present group theoretical theory. In the former case, we will discuss what kinds of anisotropic pairing states can emerge near quadrupole ordered phases. In the latter, we will point out the possibility of anisotropic pairs mediated by local fluctuations.
Finally, in Sec.\,\ref{sec:4}, we will summarize the present study.

\section{Classification of superconducting order parameters} \label{sec:2}
In this section, we explain how to classify superconducting order parameters in multi-orbital systems. Our main interest is to extend the classification of unconventional superconductivity~\cite{volovik1,volovik2,ueda1,sigrist1} into generic multi-orbital systems. Generally, the conventional BCS superconducting state is characterized by the presence of Cooper pairs with zero total momentum and the breaking of U(1) gauge symmetry.  Unconventional superconductivity additionally breaks other symmetries, for example, point group symmetry of a given system.

In this paper, for simplicity, we restrict ourselves to symmorphic-lattice systems with spacial inversion (SI) and time-reversal (TR) symmetries. In this case, superconducting order parameters, i.e., the Cooper pair wave functions can be classified by irreducible representations (IRs) of a point group $\rm P$ (See Appendix \ref{app:0}).
Furthermore, one-particle states possess the Kramers degeneracy, which can be labeled by a {\it pseudo}-spin 1/2 at each $\bm k$ point. 

In the previous studies~\cite{volovik1,volovik2,ueda1,sigrist1}, it was implicitly supposed that the transformation property of the {\it pseudo}-spin 1/2 equals to that of {\it pure}-spin 1/2. However, it is unclear whether such hypothesis holds or not in real materials, since there are orbital degrees of freedom and SOC is not negligible. Instead, we explicitly describe the transformation property of the Kramers degeneracy for the local orbital bases in multi-orbital systems, not for the band-diagonal bases. Since the classification of superconducting order parameters is very similar to that of localized multipole moment~\cite{shiina1}, we call the classified multi-orbital superconductivity ``multipole'' superconductivity. In what follows, we will show several definitions and transformation rules, and then, summarize the consequences in several tables. Through out this section, we will discuss pair amplitudes rather than the gap functions since the gap functions are readily calculated from the pair amplitudes and the symmetry properties are identical (See Appendix \ref{app:0}).

\subsection{Pair amplitude}
First of all, let us introduce an electron creation operator $c_{\ell\alpha}^\dag(\br)$ with the orbital $\ell$ and the spin $\alpha$ at the site $\br$. From a viewpoint of the classification, it is convenient to consider that $\ell$ indicates a basis function labeled by an IR of a given point group $\rm P$, and $\alpha$ denotes the Kramers degrees of freedom rather than {\it pure}-spin 1/2. See Appendix A2 for the case containing two or more atoms in a unit cell. One-particle part of Hamiltonian is diagonalized by a unitary matrix $u_{\ell\alpha, n\sigma}(\bk)$ with the band $n$, the {\it pseudo}-spin $\sigma$ and the wavenumber $\bk$. A band-based creation operator $\tilde{c}_{n\sigma}^\dag(\bk)$ is given by
\begin{subequations}\label{eq:umat}
\begin{align}
\tilde{c}_{n\sigma}^\dag(\bk)&=\frac{1}{\sqrt{N}}\sum_\br \sum_{\ell\alpha}c_{\ell\alpha}^\dag(\br) \exp[i\bk\cdot\br] u_{\ell\alpha,n\sigma}(\bk),\\
&\equiv \sum_{\ell\alpha}c_{\ell\alpha}^\dag({\bk}) u_{\ell\alpha,n\sigma}(\bk), \label{eq:cfou}
\end{align}
\end{subequations}
where $N$ is the number of unit cells. The corresponding annihilation operator is obtained by the Hermite conjugate of Eq.\,\eqref{eq:umat}. 

In the orbital bases, a pair amplitude is defined as
\begin{align}
F_{\ell\alpha,\ell'\alpha'}(\bk)\equiv \langle c_{\ell\alpha}(\bk) c_{\ell'\alpha'}(-\bk)\rangle, \label{eq:fdef}
\end{align}
where $\langle\cdot\rangle$ denotes the thermal average, and the fermion antisymmetry requires 
\begin{align}
F_{\ell\alpha,\ell'\alpha'}(\bk)=-F_{\ell'\alpha',\ell\alpha}(-\bk). \label{eq:anti}
\end{align}
Hereafter, we will discuss the classification of $F_{\ell\alpha,\ell'\alpha'}(\bk)$. 

\subsection{List of irreducible representations for the Kramers sector}\label{sec:2b}
We perform the classification of the pair amplitude $F_{\ell\alpha,\ell'\alpha'}(\bk)$ in typical point groups $\rm O$, $\rm D_4$,  and $\rm D_6$. The classification consists of that in the orbital sectors $\ell\ell'$, the Kramers sector $\alpha\alpha'$, and the wavenumber $\bk$. Once the orbital sectors are fixed, we can decompose $F$ as, 
\begin{align}
F_{\ell\alpha,\ell'\alpha'}(\bk)=\Big[\Big(\Phi_{\ell\ell'}(\bk)\sigma^0\!+{\bm d}_{\ell\ell'}(\bk)\cdot {\boldsymbol{\sigma}} \Big) i\sigma^y \Big]_{\alpha\alpha'}, 
\label{eq:F}
\end{align}
where $\sigma^0$ is a $2\times 2$ identity matrix, and $\boldsymbol{\sigma}=(\sigma^x,\sigma^y,\sigma^z)$ are the Pauli matrices in the Kramers sector. The explicit form of the Kramers pairs ($\alpha=\pm$) in each point group $\rm P$ is listed in Appendix \ref{app:2}. From the transformation property under the point group operations, we classify the Kramers part,
\begin{align}
\bar{\sigma}^{\mu}\equiv \sigma^{\mu}i\sigma^y,~(\mu=0,x,y, {\rm and\ } z)
\end{align}
into the corresponding IRs. The results are summarized in Tables \ref{table1-O}-\ref{table1-D6}. It should be noted that the generalized $\bm d$-vector, ${\bm d}_{\ell\ell'}(\bk)$, is no longer a net spin moment of Cooper pairs, although we conventionally use the unit vectors $\bm x$, $\bm y$, and $\bm z$. 

Finally, the classification of $F_{\ell\alpha,\ell'\alpha'}(\bk)$ is completed by classifying $\bk$ dependence of the basis functions, $\Phi_{\ell\ell'}(\bk)$ and ${\bm d}_{\ell\ell'}(\bk)$. Representative examples of these basis functions are listed in a column $\phi^{\varGamma}(\bk)$ in Tables \ref{table1-O}-\ref{table1-D6}. 
In SI invariant systems, all IRs are classified into even/odd parity, which is conventionally labeled with $g/u$. By adding the label $g/u$ to $\varGamma$ in an appropriate manner, one can make tables for $\rm O_h$, $\rm D_{4h}$, and $\rm D_{6h}$ groups straightforwardly. For complete set of basis functions, $\phi^{\varGamma}(\bk)$, see Ref.\,\cite{Yip}.

\begin{table}[t!]
\caption{Basis functions of IRs in O group. $\bar{\sigma}^\mu=i\sigma^\mu\sigma^y$ is represented by $\bmu=\bo, \bx, \by, \bz$, symbolically. Index $a(b)$ of $\bmu_{a(b)}$ represents that the pair consists of one of the non-Kramers doublet $a(b)$ in $\varGamma_8$ (Appendix \ref{app:2}) and the other orbital $\varGamma_6$ or $\varGamma_7$. $\bmu_{a\pm}=-\frac{1}{2}(\bmu_a\pm\sqrt{3}\bmu_b)$ and $\bmu_{b\pm}=\frac{1}{2}(-\bmu_b\pm\sqrt{3}\bmu_a)$. $\tau^\mu$'s are the Pauli matrices in the orbital space spanned by the non-Kramers degrees of freedom ($a/b$). ${\bm \zeta}=\cos\theta(\tau^0,\tau^0,\tau^0)+\sin\theta(\tau^z_{-},\tau^z_{+},\tau^z)$ and ${\bm \eta}=\cos\theta(\tau^y,\tau^y,\tau^y)+i\sin\theta(\tau^x_{-},\tau^x_{+},\tau^x)$, where $\tau^z_{\pm}=-\frac{1}{2}(\tau^z\pm\sqrt{3}\tau^x)$ and $\tau^x_{\pm}=\frac{1}{2}(-\tau^x\pm\sqrt{3}\tau^z)$. $\theta$ is an arbitraly real parameter.}
\label{table1-O}
\begin{tabular*}{84mm}{>{\centering\arraybackslash}p{5mm}>{\centering\arraybackslash}p{30mm}c>{\centering\arraybackslash}p{21mm}}\hline\hline
IR & $\phi^\varGamma(\bk)$ & $\varGamma_6\otimes\varGamma_6$ / $\varGamma_7\otimes\varGamma_7$ & $\varGamma_6\otimes\varGamma_7$ \\ \hline
$A_1$ & $k_x^2+k_y^2+k_z^2$& $\bo$ &  \\
$A_2$ & $k_xk_yk_z$        &  & $\bo$ \\
$E$   & $(3k_z^2-k^2,k_x^2-k_y^2)$ &  & \\
$T_1$ & $(k_x,k_y,k_z)$    & $(\bx,\by,\bz)$ & \\
$T_2$ & $(k_yk_z,k_zk_x,k_xk_y)$ & & $(\bx,\by,\bz)$ \\
\hline
\end{tabular*}\vspace{.3mm}

\begin{tabular*}{84mm}{>{\centering\arraybackslash}p{5mm}>{\centering\arraybackslash}p{26mm}>{\centering\arraybackslash}p{21mm}>{\centering\arraybackslash}p{30mm}}  \hline
IR & $\varGamma_6\otimes\varGamma_8$ & $\varGamma_7\otimes\varGamma_8$ & $\varGamma_8\otimes\varGamma_8$ \\ \hline 
$A_1$ &&& $\tau^0\bo$ \\
$A_2$ &&& $\tau^y\bo$ \\
$E$   & $(\bo_b,\bo_a)$ & $(\bo_a,-\bo_b)$ & ($\tau^z,\tau^x)\bo$ \\
$T_1$ & $(\bx_{b+},\by_{b-},\bz_b)$ & $(\bx_{a+},\by_{a-},\bz_a)$ & $(\zeta^1\bx,\zeta^2\by,\zeta^3\bz)$ \\
$T_2$ & $(\bx_{a+},\by_{a-},\bz_a)$ & $(\bx_{b+},\by_{b-},\bz_b)$ & $(\eta^1\bx,\eta^2\by,\eta^3\bz)$ \\
\hline \hline
\end{tabular*}
\caption{Basis functions of IRs in D$_4$ group.}
\label{table1-D4}
\begin{tabular*}{84mm}{>{\centering\arraybackslash}p{5mm}>{\centering\arraybackslash}p{28mm}c>{\centering\arraybackslash}p{22mm}} \hline \hline
IR & $\phi^\varGamma(\bk)$ & $\varGamma_6\otimes\varGamma_6$ / $\varGamma_7\otimes\varGamma_7$ & $\varGamma_6\otimes\varGamma_7$ \\ \hline
$A_1$ & $k_z^2$     & $\bo$ &  \\
$A_2$ & $k_z$       & $\bz$ &  \\
$B_1$ & $k_x^2-k_y^2$ &  & $\bo$ \\
$B_2$ & $k_xk_y$      &  & $\bz$ \\
$E$   & $(k_x,k_y)$   & $(\bx,\by)$ & $(\bx,-\by)$ \\
\hline \hline
\end{tabular*}
\caption{Basis functions of IRs in D$_6$ group. $i=7 (8)$ corresponds to upper(lower) expressions.}
\label{table1-D6}
\begin{tabular*}{84mm}{>{\centering\arraybackslash}p{5mm}>{\centering\arraybackslash}p{22mm}>{\centering\arraybackslash}p{12mm}>{\centering\arraybackslash}p{12mm}>{\centering\arraybackslash}p{12mm}>{\centering\arraybackslash}p{12mm}}  \hline \hline
IR & $\phi^\varGamma(\bk)$ & $\varGamma_i\otimes\varGamma_i$ & $\varGamma_9\otimes\varGamma_9$ & $\varGamma_7\otimes\varGamma_8$ & $\varGamma_i\otimes\varGamma_9$ \\ \hline
$A_1$ & $k_z^2$     & $\bo$ & $\bo$ & &\\
$A_2$ & $k_z$       & $\bz$ & $\bz$ & &\\
$B_1$ & $k_y^3-3k_yk_x^2$ &  & $\by$ & $\by$ &\\
$B_2$ & $k_x^3-3k_xk_y^2$ &  & $\bx$ & $\bx$ &\\
$E_1$ & $(k_x,k_y)$   & $(\bx,\pm\by)$ & & & $(\bx,\mp\by)$ \\
$E_2$ & $(2k_xk_y,k_x^2-k_y^2)$ & & & $(i\bz,\bo)$ & $(i\bz,\mp\bo)$ \\
\hline \hline
\end{tabular*}
\end{table}

Now, we discuss the consequence of the lists in Tables \ref{table1-O}-\ref{table1-D6}. We realize that even in a single-orbital system, orbital character can play crucial roles. Within the whole 32 point groups, there exists {\it one and only one} nontrivial combination whose transformation properties are completely different from the other cases. That is $\varGamma_9\otimes\varGamma_9$ in D$_6$ and the equivalent groups, which do not include $E_1$ representation in sharp contrast to the other products $\varGamma_7\otimes\varGamma_7$ or $\varGamma_8\otimes\varGamma_8$. In this case, the gap functions can show an anomalous $\bk$ dependence, which explains the emergence of an exotic gap structure in the microscopic study for UPt$_3$~\cite{UPt3}. To the best of our knowledge, this point has not been recognized in the long history of research in superconductivity, which is one of nontrivial results in this study. 

As highlighted in $\varGamma_9\otimes\varGamma_9$ in $\rm D_6$ point group, it is noteworthy that, in Tables \ref{table1-O}-\ref{table1-D6}, the Kramers sector takes different IRs, depending on the constituting orbitals. For example, direct products for {\it pure}-spin 1/2 s-orbital electrons in $\rm D_{4h}$ point group, which correspond to $\varGamma_{6g}\otimes\varGamma_{6g}$ in Table \ref{table1-D4}, include $A_{1g}$ and $A_{2g}$ representations. In contrast, $\varGamma_{6g}\otimes\varGamma_{7g}$ includes $B_{1g}$ and $B_{2g}$, while it does not include $A_{1g}$ and $A_{2g}$ representations. Moreover, a spin-singlet state~\cite{memo0} described by $\bm 0$ in $\varGamma_{6g}\otimes\varGamma_{7g}$ belongs to $B_{1g}$, while that in $\varGamma_{6g}\otimes\varGamma_{6g}$ belongs to the identity representation. This is an essential aspect of the electron pairing in multi-orbital systems.

Note that in Table \ref{table1-O}, the pairs including non-Kramers doublet $\varGamma_8$ are complicated because the $\varGamma_8$ bases labeled by $a$ and $b$ (See Appendix \ref{app:2}) are inseparable under the point group operations. This degeneracy also can lead to the exotic pairing state, as recently proposed for the superconductivity in half-Heusler semimetal YPtBi~\cite{HKim,Brydon,memo1}. About the inter-orbital pairs including $\varGamma_8$ states, the classification can be performed by introducing the Pauli matrices $\tau^\mu_{a(b)}$ acting on $\varGamma_{8a(b)}$ and $\varGamma_{6,7}$.

\begin{table*}[h!t]
\begin{center}
\caption{Basis functions of IRs in O group. The following abbreviations are used; 
$\phi^\varGamma_i=\phi^\varGamma_i(\bk)$, 
$\phi^E_{1\pm}=\frac{1}{2}(-\phi^E_1\pm\sqrt{3}\phi^E_2)$, 
$\phi^E_{2\pm}=-\frac{1}{2}(\phi^E_2\pm\sqrt{3}\phi^E_1)$, and 
$\phi^{T_1}_i(\bk)=k_i$, $\phi^{T_2}(\bk)=\tk_i$ with $i=1,2,3$. 
Basis fucntions in $\varGamma_6\otimes\varGamma_8$ space are obtained by replacing ${\boldsymbol{\mu}}_a\to{\boldsymbol{\mu}}_b, {\boldsymbol{\mu}}_b\to-{\boldsymbol{\mu}}_a$ with ${\boldsymbol{\mu}}={\bo, \bx, \by, \bz}$ in the table of $\varGamma_7\otimes\varGamma_8$ space.  The other notations are the same as in Table \ref{table1-O}.}
\label{table2-O}
\begin{tabular}{>{\centering\arraybackslash}p{15mm}>{\centering\arraybackslash}p{50mm}>{\centering\arraybackslash}p{105mm}}
\hline\hline
IR & \multicolumn{2}{c}{$\varGamma_6\otimes\varGamma_6$ / $\varGamma_7\otimes\varGamma_7$} \\ \hline
$A_{1}$ &$\phi^{A_1}\bo$ &$k_1\bx\!+\!k_2\by\!+\!k_3\bz$\\
$A_{2}$ &$\phi^{A_2}\bo$ &$\tk_1\bx\!+\!\tk_2\by\!+\!\tk_3\bz$\\
$E$ &$(\phi^E_1,\phi^E_2)\bo$
&$\big( \frac{k_1}{\sqrt{3}}\bx\!+\!\frac{k_2}{\sqrt{3}}\by\!-\!\frac{2k_3}{\sqrt{3}}\bz,  k_2\by\!-\!k_1\bx\big),
~\big( \tk_1\bx\!-\!\tk_2\by, \frac{\tk_1}{\sqrt{3}}\bx\!+\!\frac{\tk_2}{\sqrt{3}}\by\!-\!\frac{2\tk_3}{\sqrt{3}}\bz\big)$ \\
$T_{1}$ &$(k_1,k_2,k_3)\bo$
&$(\phi^{A_1}\bx,\phi^{A_1}\by,\phi^{A_1}\bz),
~(k_2\bz\!-\!k_3\by,k_3\bx\!-\!k_1\bz,k_1\by\!-\!k_2\bx),$ \\
&& $(\phi^E_{1+}\bx,\phi^E_{1-}\by,\phi^E_1\bz),
~(\tk_2\bz\!+\!\tk_3\by,\tk_3\bx\!+\!\tk_1\bz,\tk_1\by\!+\!\tk_2\bx)$ \\
$T_{2}$ &$(\tk_1,\tk_2,\tk_3)\bo$
& $(\phi^{A_2}\bx,\phi^{A_2}\by,\phi^{A_2}\bz),
~(k_2\bz\!+\!k_3\by,k_3\bx\!+\!k_1\bz,k_1\by\!+\!k_2\bx)$, \\
&& $(\phi^E_{2+}\bx,\phi^E_{2-}\by,\phi^E_2\bz),
~(\tk_2\bz\!-\!\tk_3\by,\tk_3\bx\!-\!\tk_1\bz,\tk_1\by\!-\!\tk_2\bx)$ \\ \hline\hline
IR & \multicolumn{2}{c}{$\varGamma_6\otimes\varGamma_7$} \\ \hline
$A_{1}$ &$\phi^{A_2}\bo$ &$\tk_1\bx\!+\!\tk_2\by\!+\!\tk_3\bz$\\ 
$A_{2}$ &$\phi^{A_1}\bo$ & $k_1\bx\!+\!k_2\by\!+\!k_3\bz$ \\ 
$E$ &$(\phi^E_2,-\phi^E_1)\bo$
& $\big( k_1\bx \!-\! k_2\by, \frac{k_1}{\sqrt{3}}\bx\!+\!\frac{k_2}{\sqrt{3}}\by\!-\!\frac{2k_3}{\sqrt{3}}\bz\big),
~\big( \frac{\tk_1}{\sqrt{3}}\bx\!+\!\frac{\tk_2}{\sqrt{3}}\by\!-\!\frac{2\tk_3}{\sqrt{3}}\bz,  \tk_2\by\!-\!\tk_1\bx\big)$\\
$T_{1}$ &$(\tk_1,\tk_2,\tk_3)\bo$
& $(\phi^{A_2}\bx,\phi^{A_2}\by,\phi^{A_2}\bz),
~(k_2\bz\!+\!k_3\by,k_3\bx\!+\!k_1\bz,k_1\by\!+\!k_2\bx)$, \\
&& $(\phi^E_{2+}\bx,\phi^E_{2-}\by,\phi^E_2\bz),
~(\tk_2\bz\!-\!\tk_3\by,\tk_3\bx\!-\!\tk_1\bz,\tk_1\by\!-\!\tk_2\bx)$\\
$T_{2}$ &$(k_1,k_2,k_3)\bo$
& $(\phi^{A_1}\bx,\phi^{A_1}\by,\phi^{A_1}\bz),
~(k_2\bz\!-\!k_3\by,k_3\bx\!-\!k_1\bz,k_1\by\!-\!k_2\bx)$, \\ 
&& $(\phi^E_{1+}\bx,\phi^E_{1-}\by,\phi^E_1\bz),
~(\tk_2\bz\!+\!\tk_3\by,\tk_3\bx\!+\!\tk_1\bz,\tk_1\by\!+\!\tk_2\bx)$\\ 
\hline\hline
IR & \multicolumn{2}{c}{$\varGamma_7\otimes\varGamma_8$ / $\varGamma_6\otimes\varGamma_8~~~({\boldsymbol{\mu}}_a\to{\boldsymbol{\mu}}_b, {\boldsymbol{\mu}}_b\to-{\boldsymbol{\mu}}_a$)} \\ \hline
$A_{1}$ &$\phi^E_1\bo_a\!-\!\phi^E_2\bo_b$
& $\big\{k_1\bx_{a+}\!+\!k_2\by_{a-}\!+\!k_3\bz_a,~(k\to \tk,a\to b)\big\}$ \\
$A_{2}$ &$\phi^E_2\bo_a\!+\!\phi^E_1\bo_b$
& $\big\{\tk_1\bx_{a+}\!+\!\tk_2\by_{a-}\!+\!\tk_3\bz_a,~(\tk\to k,a\to b)\big\}$ \\
$E$ &$(\phi^{A_1}\bo_a,-\phi^{A_1}\bo_b),~(\phi^{A_2}\bo_b,\phi^{A_2}\bo_a),$& 
$\big\{\big( \frac{k_1}{\sqrt{3}}\bx_{a+}\!+\!\frac{k_2}{\sqrt{3}}\by_{a-}\!-\!\frac{2k_3}{\sqrt{3}}\bz_a,  k_2\by_{a-}\!-\!k_1\bx_{a+}\big),~
(k\!\to\! \tk, a \to b)\big\}$,\\
&$(\phi^E_1\bo_a\!+\!\phi^E_2\bo_b,-\phi^E_2\bo_a\!+\!\phi^E_1\bo_b)$& $\big\{\big( \tk_1\bx_{a+}\!-\!\tk_2\by_{a-}, \frac{\tk_1}{\sqrt{3}}\bx_{a+}\!+\!\frac{\tk_2}{\sqrt{3}}\by_{a-}\!-\!\frac{2\tk_3}{\sqrt{3}}\bz_a\big),~
(\tk\!\to\! k, a \to b)\big\}$\\
$T_{1}$ &$(k_1\bo_{a+},k_2\bo_{a-},k_3\bo_a)$,& 
$\big\{ \phi^{A_1}(\bx_{a+},\by_{a-},\bz_a),~
(A_1\to A_2, a\to b)\big\}$,\\
&$(\tk_1\bo_{b+},\tk_2\bo_{b-},\tk_3\bo_b)$& 
$\big\{(\phi^E_{1+}\bx_{a+},\phi^E_{1-}\by_{a-},\phi^E_1\bz_a),~ (\phi^E_1\to\phi^E_2, a\to b)\big\}$,\\
&&$
\big\{(k_2\bz_a\!-\!k_3\by_{a-},k_3\bx_{a+}\!-\!k_1\bz_a,k_1\by_{a-}\!-\!k_2\bx_{a+}),~
(k\!\to\! \tk, a\!\to\!b)\big\}$,\\
&&
$
\big\{(\tk_2\bz_a\!+\!\tk_3\by_{a-},\tk_3\bx_{a+}\!+\!\tk_1\bz_a,\tk_1\by_{a-}\!+\!\tk_2\bx_{a+}),~
(\tk\!\to\! k, a\!\to\!b)\big\}
$ \\
$T_{2}$ &$(\tk_1\bo_{a+},\tk_2\bo_{a-},\tk_3\bo_a)$,
& $\big\{\phi^{A_1}(\bx_{b+},\by_{b-},\bz_{b}),~(A_1\to A_2, b\to a)\big\}$,\\
&$(k_1\bo_{b+},k_2\bo_{b-},k_3\bo_b)$& 
$\big\{(\phi^E_{1+}\bx_{b+},\phi^E_{1-}\by_{b-},\phi^E_1\bz_b),~(\phi^E_1\to \phi^E_2, b\to a)\big\}$ \\
&&
$
\big\{(k_2\bz_a\!+\!k_3\by_{a-},k_3\bx_{a+}\!+\!k_1\bz_a,k_1\by_{a-}\!+\!k_2\bx_{a+}),~(k\!\to\! \tk, a\!\to\!b)\big\}$,\\
&&
$
\big\{(\tk_2\bz_a\!-\!\tk_3\by_{a-},\tk_3\bx_{a+}\!-\!\tk_1\bz_a,\tk_1\by_{a-}\!-\!\tk_2\bx_{a+}),~
(\tk\!\to\! k, a\!\to\!b)\big\}
$ \\ \hline \hline
IR & \multicolumn{2}{c}{$\varGamma_8\otimes\varGamma_8$} \\ \hline 
$A_{1}$ & $\phi^{A_1}\tau^0\bo,~\phi^E_{1}\tau^z\bo\!+\!\phi^E_{2}\tau^x\bo$,~$\phi^{A_2}\tau^y\bo$ 
& $\big\{k_1\zeta^1\bx\!+\!k_2\zeta^2\by\!+\!k_3\zeta^3\bz,~(k\to \tk,\zeta\to \eta)\big\}
$ \\
$A_{2}$ & $\phi^{A_2}\tau^0\bo,~\phi^E_{2}\tau^z\bo\!-\!\phi^E_{1}\tau^x\bo$,~$\phi^{A_1}\tau^y\bo$
&$\big\{\tk_1\zeta^1\bx\!+\!\tk_2\zeta^2\by\!+\!\tk_3\zeta^3\bz,~(\tk\to k,\zeta\to \eta)\big\}
$ \\ 
$E$&$(\phi^{A_1}\tau^z,\phi^{A_1}\tau^x)\bo,(\phi^{A_2}\tau^x,-\phi^{A_2}\tau^z)\bo,$
&$\big\{\big( \frac{k_1}{\sqrt{3}}\zeta^1\bx\!+\!\frac{k_2}{\sqrt{3}}\zeta^2\by\!-\!\frac{2k_3}{\sqrt{3}}\zeta^3\bz,  k_2\zeta^2\by\!-\!k_1\zeta^1\bx\big),~(k\to \tk, \zeta \to \eta)\big\}$,\\
&$(\phi^E_1\tau^z\!-\!\phi^E_2\tau^x,-\phi^E_2\tau^z\!-\!\phi^E_1\tau^x)\bo,$&
$\big\{\big( \tk_1\zeta^1\bx \!+\! \tk_2\zeta^2\by, \frac{\tk_1}{\sqrt{3}}\zeta^1\bx\!+\!\frac{\tk_2}{\sqrt{3}}\zeta^2\by\!+\!\frac{2\tk_3}{\sqrt{3}}\zeta^3\bz\big),~ (\tk\to k, \zeta\to\eta)\big\}$\\
&$(\phi^E_1,\phi^E_2)\tau^0\bo$,~$(\phi^E_2,-\phi^E_1)\tau^y\bo$&\\
$T_1$ &$(k_1,\,k_2,\,k_3)\tau^0\bo$,
&$\big\{\phi^{A_1}(\zeta^1\bx,\,\zeta^2\by,\,\zeta^3\bz),~
(A_1\to A_2, \zeta\to\eta)\big\}
,$\\
&$(\tk_1,\,\tk_2,\,\tk_3)\tau^y\bo$,
&$\big\{(\phi^E_{1+}\zeta^1\bx,\,\phi^E_{1-}\zeta^2\by,\,\phi^E_1\zeta^3\bz),~
(\phi^E_1\to\phi^E_2, \zeta\to\eta)\big\}
,$\\
&$(k_1\tau^z_-,\,k_2\tau^z_+,\,k_3\tau^z)\bo$,
&$\big\{(k_2\zeta^3\bz\!-\!k_3\zeta^2\by,\,k_3\zeta^1\bx\!-\!k_1\zeta^3\bz,\,k_1\zeta^2\by\!-\!k_2\zeta^1\bx), ~(k\to \tk,\zeta\to \eta)\big\}$,\\
&$(\tk_1\tau^x_-,\,\tk_2\tau^x_+,\,\tk_3\tau^x)\bo$
&$\big\{(\tk_2\zeta^3\bz\!+\!\tk_3\zeta^2\by,\,\tk_3\zeta^1\bx\!+\!\tk_1\zeta^3\bz,\,\tk_1\zeta^2\by\!+\!\tk_2\zeta^1\bx), ~(\tk\to k,\zeta\to \eta)\big\}$\\
$T_2$ &$(\tk_1,\,\tk_2,\,\tk_3)\tau^0\bo$,
&$\big\{\phi^{A_2}(\zeta^1\bx,\,\zeta^2\by,\,\zeta^3\bz),~(A_2\to A_1, \zeta\to\eta)\big\}
,$\\
&$(k_1,\,k_2,\,k_3)\tau^y\bo$,
&$(\phi^E_{2+}\zeta^1\bx,\,\phi^E_{2-}\zeta^2\by,\,\phi^E_2\zeta^3\bz),
~(\phi^E_2\to\phi^E_1, \zeta\to\eta)\big\}
,$\\
&$(\tk_1\tau^z_-,\,\tk_2\tau^z_+,\,\tk_3\tau^z)\bo,$
&$\big\{(\tk_2\zeta^3\bz\!-\!\tk_3\zeta^2\by,\, \tk_3\zeta^1\bx\!-\!\tk_1\zeta^3\bz,\, \tk_1\zeta^2\by\!-\!\tk_2\zeta^1\bx), ~(\tk\to k,\zeta\to \eta)\big\}$,\\
&$(k_1\tau^x_-,\,k_2\tau^x_+,\,k_3\tau^x)\bo$
&$\big\{(k_2\zeta^3\bz\!+\!k_3\zeta^2\by,\, k_3\zeta^1\bx\!+\!k_1\zeta^3\bz,\, k_1\zeta^2\by\!+\!k_2\zeta^1\bx), ~(k\to \tk,\zeta\to \eta)\big\}$\\  \hline\hline
\end{tabular}
\end{center}
\end{table*}

\subsection{List of full irreducible representations}\label{sec:2c}

Now, let us complete a list of IRs of gap functions, which is constructed via the subduction of 
\begin{align}
\big(\bk\ {\rm dependence\ }\phi^{\varGamma}(\bk)\big)\otimes ({\rm Kramers\ part}) \downarrow {\rm P}, \label{eq:subduc}
\end{align}
(See Appendix \ref{app:0}).
The results are summarized in Tables \ref{table2-O}-\ref{table2-D6}. These basis functions obtained by the subduction should still be antisymmetrized to meet the fermion antisymmetry. For this purpose, it is instructive to explicitly write down the pair amplitudes of Eq.\,\eqref{eq:F} as,
\begin{align}
F_{\ell\alpha,\ell'\alpha'}(\bk) = \sum_{\mu\nu} d^{\mu\nu}(\bk) \tau^\nu_{\ell\ell'}\bar{\sigma}^\mu_{\alpha\alpha'},  \label{eq:Flm}
\end{align}
where the matrix $\tau^\nu_{\ell\ell'}$ characterizes the orbital sector of the pair amplitudes. In the followings, we call $\tau^\nu_{\ell\ell'}\bar{\sigma}^\mu_{\alpha\alpha'}$ in Eq.\,\eqref{eq:Flm} a multipole part of the pair amplitudes and denote $\boldsymbol{\tau\bar\sigma}$ symbolically. In terms of $d^{\mu\nu}(\bk),\Phi_{\ell\ell'}(\bk)$ and $d_{\ell\ell'}^\mu(\bk)$ in Eq.\,\eqref{eq:F} are given by,
\begin{subequations}
\begin{align}
\Phi_{\ell\ell'}(\bk) &=\sum_{\nu}d^{0\nu}(\bk)\tau^\nu_{\ell\ell'}, \\
d^\mu_{\ell\ell'}(\bk) &=\sum_{\nu}d^{\mu\nu}(\bk)\tau^\nu_{\ell\ell'}.
\end{align}
\end{subequations}
The size of matrix $\tau^\nu_{\ell\ell'}$ depends on a given number of orbitals. For example, $\tau^\nu_{\ell\ell'}$ is the Gell-Mann matrix in three-orbital systems with $\varGamma_6\otimes\varGamma_8$ and $\varGamma_7\otimes\varGamma_8$ in O group, otherwise the Pauli matrix in two-orbital systems. Hereafter, let us consider two-orbital systems for simplicity. The generalization to generic multi-orbital systems is straightforward. For the $\boldsymbol{\tau\bar\sigma}$ pairing states, we can define orbital ($o$) singlet/triplet after spin ($s$) singlet/triplet. In what follows $o$-triplet $s$-singlet or $o$-singlet $s$-triplet is referred to be multipole ($m$) singlet, while $o$-singlet $s$-singlet or $o$-triplet $s$-triplet to be $m$-triplet. Note that the singlet(triplet) just means odd(even) under the exchange of the corresponding indices.

\begin{table}[t!]
\caption{Basis functions of IRs in D$_4$ group.}
\label{table2-D4}
\begin{tabular}{>{\centering\arraybackslash}p{8mm}>{\centering\arraybackslash}p{2cm} >{\centering\arraybackslash}p{5.5cm}}
\hline\hline
IR & \multicolumn2{c}{$\varGamma_6\otimes\varGamma_6$ / $\varGamma_7\otimes\varGamma_7$} \\ \hline
$A_1$ & $\phi^{A_1}\bo$ & $\phi^{A_2}\bz,~ \phi^E_1\bx+\phi^E_2\by$ \\
$A_2$ & $\phi^{A_2}\bo$ & $\phi^{A_1}\bz,~ \phi^E_2\bx-\phi^E_1\by$ \\
$B_1$ & $\phi^{B_1}\bo$ & $\phi^{B_2}\bz,~ \phi^E_1\bx-\phi^E_2\by$ \\
$B_2$ & $\phi^{B_2}\bo$ & $\phi^{B_1}\bz,~ \phi^E_2\bx+\phi^E_1\by$ \\
$E$ & $(\phi^E_1,\phi^E_2)\bo$ & $\phi^{A_1}(\bx,\by),\phi^{A_2}(\by,-\bx)$, \\
& & $\phi^{B_1}(\bx,-\by), \phi^{B_2}(\by,\bx),$ \\
& & $(\phi^E_2,-\phi^E_1)\bz$ \\ \hline \hline
\end{tabular}

\begin{tabular}{>{\centering\arraybackslash}p{8mm}>{\centering\arraybackslash}p{2cm} >{\centering\arraybackslash}p{5.5cm}}
IR & \multicolumn2{c}{$\varGamma_6\otimes\varGamma_7$} \\ \hline
$A_1$ & $\phi^{B_1}\bo$ & $\phi^{B_2}\bz,~ \phi^E_1\bx-\phi^E_2\by$ \\
$A_2$ & $\phi^{B_2}\bo$ & $\phi^{B_1}\bz,~ \phi^E_2\bx+\phi^E_1\by$ \\
$B_1$ & $\phi^{A_1}\bo$ & $\phi^{A_2}\bz,~ \phi^E_1\bx+\phi^E_2\by$ \\
$B_2$ & $\phi^{A_2}\bo$ & $\phi^{A_1}\bz,~ \phi^E_2\bx-\phi^E_1\by$ \\
$E$ & $(\phi^E_1,-\phi^E_2)\bo$ & $\phi^{A_1}(\bx,-\by),~ \phi^{A_2}(\by,\bx)$,\\
& & $\phi^{B_1}(\bx,\by),~ \phi^{B_2}(\by,-\bx),$ \\
& & $(\phi^E_2,\phi^E_1)\bz$ \\  \hline \hline
\end{tabular}
\end{table}
\begin{table}[t!]
\caption{Basis functions of IRs in D$_6$ group. Expressions for $\varGamma_{7(8)}$ correspond to upper(lower) signs.}
\label{table2-D6}

\begin{tabular}{>{\centering\arraybackslash}p{8mm}>{\centering\arraybackslash}p{2cm} >{\centering\arraybackslash}p{5.5cm}}
\hline
\hline
IR & \multicolumn{2}{c}{$\varGamma_7\otimes\varGamma_7$(upper) / $\varGamma_8\otimes\varGamma_8$(lower)} \\ \hline
$A_1$ &$\phi^{A_1}\bo$& $\phi^{A_2}\bz,~ \phi^{E_1}_1\bx\pm \phi^{E_1}_2\by$ \\
$A_2$ &$\phi^{A_2}\bo$& $\phi^{A_1}\bz,~ \phi^{E_1}_2\bx\mp \phi^{E_1}_1\by$ \\
$B_1$ &$\phi^{B_1}\bo$& $\phi^{B_2}\bz,~ \phi^{E_2}_1\bx\pm \phi^{E_2}_2\by$ \\
$B_2$ &$\phi^{B_2}\bo$& $\phi^{B_1}\bz,~ \phi^{E_2}_2\bx\mp \phi^{E_2}_1\by$ \\
$E_1$ &$(\phi^{E_1}_1,\phi^{E_1}_2)\bo$& $\phi^{A_1}(\bx,\pm \by),~ \phi^{A_2}(\by,\mp \bx),$ \\ 
&& $(\phi^{E_2}_2\bx\pm \phi^{E_2}_1\by,\phi^{E_2}_1\bx\mp \phi^{E_2}_2\by),$ \\ 
&& $(\phi^{E_1}_2,-\phi^{E_1}_1)\bz$ \\ 
$E_2$ &$(\phi^{E_2}_1,\phi^{E_2}_2)\bo$& $\phi^{B_1}(\bx,\pm \by),~ \phi^{B_2}(\by,\mp \bx),$ \\ 
&& $(\phi^{E_1}_2\bx\pm \phi^{E_1}_1\by,\phi^{E_1}_1\bx\mp \phi^{E_1}_2\by),$ \\ 
&& $(\phi^{E_2}_2,-\phi^{E_2}_1)\bz$ \\  \hline \hline
IR & \multicolumn{2}{c}{$\varGamma_9\otimes\varGamma_9$} \\ \hline
$A_1$ &$\phi^{A_1}\bo$& $\phi^{A_2}\bz,~ \phi^{B_1}\by,~ \phi^{B_2}\bx$ \\
$A_2$ &$\phi^{A_2}\bo$& $\phi^{A_1}\bz,~ \phi^{B_2}\by,~ \phi^{B_1}\bx$ \\
$B_1$ &$\phi^{B_1}\bo$& $\phi^{B_2}\bz,~ \phi^{A_1}\by,~ \phi^{A_2}\bx$ \\
$B_2$ &$\phi^{B_2}\bo$& $\phi^{B_1}\bz,~ \phi^{A_2}\by,~ \phi^{A_1}\bx$ \\
$E_1$ &$(\phi^{E_1}_1,\phi^{E_1}_2)\bo$& $(\phi^{E_2}_1,\phi^{E_2}_2)\by,~ (\phi^{E_2}_2,-\phi^{E_2}_1)\bx,$ \\ 
&& $(\phi^{E_1}_2,-\phi^{E_1}_1)\bz$ \\
$E_2$ &$(\phi^{E_2}_1,\phi^{E_2}_2)\bo$& $(\phi^{E_1}_1,\phi^{E_1}_2)\by,~ (\phi^{E_1}_2,-\phi^{E_1}_1)\bx,$ \\ 
&& $(\phi^{E_2}_2,-\phi^{E_2}_1)\bz$ \\ \hline \hline
\end{tabular}

\begin{tabular}{>{\centering\arraybackslash}p{8mm}>{\centering\arraybackslash}p{2.7cm} >{\centering\arraybackslash}p{4.8cm}}
IR & \multicolumn{2}{c}{$\varGamma_7\otimes\varGamma_8$} \\ \hline
$A_1$ &$\phi^{B_1}\by,~ \phi^{B_2}\bx$& $\phi^{E_2}_1\bz-i\phi^{E_2}_2\bo$ \\
$A_2$ &$\phi^{B_2}\by,~ \phi^{B_2}\bx$& $\phi^{E_2}_2\bz+i\phi^{E_2}_1\bo$ \\
$B_1$ &$\phi^{A_1}\by,~ \phi^{A_2}\bx$& $\phi^{E_1}_1\bz-i\phi^{E_1}_2\bo$ \\
$B_2$ &$\phi^{A_2}\by,~ \phi^{A_1}\bx$& $\phi^{E_1}_2\bz+i\phi^{E_1}_1\bo$ \\
$E_1$ &$(\phi^{E_2}_1,\phi^{E_2}_2)\by,$& $\phi^{B_1}(\bz,-i\bo),~ \phi^{B_2}(i\bo,\bz),$ \\
&$(\phi^{E_2}_2,-\phi^{E_2}_1)\bx$& $(\phi^{E_1}_2\bz- i\phi^{E_1}_1\bo,\phi^{E_1}_1\bz+ i\phi^{E_1}_2\bo),$ \\ 
$E_2$ &$(\phi^{E_1}_1,\phi^{E_1}_2)\by,$& $\phi^{A_1}(\bz,-i\bo),~ \phi^{A_2}(i\bo,\bz)$ \\ 
&$(\phi^{E_1}_2,-\phi^{E_1}_1)\bx$& $(\phi^{E_2}_2\bz- i\phi^{E_2}_1\bo,\phi^{E_2}_1\bz+ i\phi^{E_2}_2\bo)$ \\ \hline \hline
IR & \multicolumn{2}{c}{$\varGamma_7\otimes\varGamma_9$(upper) / $\varGamma_8\otimes\varGamma_9$(lower)} \\ \hline
$A_1$ &$\phi^{E_1}_1\bx\mp \phi^{E_1}_2\by$& $\phi^{E_2}_1\bz\pm i\phi^{E_2}_2\bo$ \\
$A_2$ &$\phi^{E_1}_2\bx\pm \phi^{E_1}_1\by$& $\phi^{E_2}_2\bz\mp i\phi^{E_2}_1\bo$ \\
$B_1$ &$\phi^{E_2}_1\bx\mp \phi^{E_2}_2\by$& $\phi^{E_1}_1\bz\pm i\phi^{E_1}_2\bo$ \\
$B_2$ &$\phi^{E_2}_2\bx\pm \phi^{E_2}_1\by$& $\phi^{E_1}_2\bz\mp i\phi^{E_1}_1\bo$ \\
$E_1$ &$\phi^{A_1}(\bx,\mp\by),$& $\phi^{B_1}(\bz,\pm i\bo),~ \phi^{B_2}(i\bo,\mp\bz),$ \\ 
&$\phi^{A_2}(\by,\pm\bx),$& $~ (\phi^{E_1}_2\bz\pm i \phi^{E_1}_1\bo,\phi^{E_1}_1\bz\mp i\phi^{E_1}_2\bo)$ \\ 
&\multicolumn{2}{l}{$(\phi^{E_2}_2\bx\mp \phi^{E_2}_1\by, \phi^{E_2}_1\bx\pm \phi^{E_2}_2\by)$}  \\ 
$E_2$ &$\phi^{B_1}(\bx,\mp\by)$,& $\phi^{A_1}(\bz,\pm i\bo),~ \phi^{A_2}(i\bo,\mp\bz)$ \\ 
&$\phi^{B_2}(\by,\pm\bx),$& $(\phi^{E_2}_2\bz\pm i \phi^{E_2}_1\bo,\phi^{E_2}_1\bz\mp i\phi^{E_2}_2\bo)$ \\ 
&\multicolumn{2}{l}{$(\phi^{E_1}_2\bx\mp \phi^{E_1}_1\by, \phi^{E_1}_1\bx\pm \phi^{E_1}_2\by)$}  \\  \hline\hline 
\end{tabular}
\end{table}

Let us discuss the properties of $d^{\mu\nu}(\bk)$. First, the fermion antisymmetry imposes a constraint,
\begin{align}\label{eq:2c-2}
d^{\mu\nu}(\bk)\tau^\nu\bar{\sigma}^\mu=-d^{\mu\nu}(-\bk)(\tau^\nu)^T(\bar{\sigma}^\mu)^T,
\end{align}
where $A^T$ denotes the transpose of the matrix $A$. 
From this relation, one can see that $d^{\mu\nu}(\bk)$ should be even (odd) under the transform $\bk\rightarrow-\bk$ for $m$-singlet (triplet) pairings. 
Next, the TR symmetry imposes another constraint,
\begin{equation}\label{eq:2c-2TR}
d^{\mu\nu}(\bk)\tau^\nu\bar{\sigma}^\mu=-d^{\mu\nu*}(-\bk)(\tau^\nu)^T(\bar{\sigma}^\mu)^T.
\end{equation}
From Eqs.\,\eqref{eq:2c-2} and \eqref{eq:2c-2TR}, we find that $d^{\mu\nu}(\bk)$ is real whenever the TR symmetry is preserved. 
Note also that the multipole part of pair amplitudes $\boldsymbol{\tau\bar\sigma}$ is 
TR even (odd) for $m$-singlet (triplet), according to the fact $(\tau^\nu)^T(\bar{\sigma}^\mu)^T=-\tau^\nu\bar{\sigma}^\mu$ for $m$-singlet and $\tau^\nu\bar{\sigma}^\mu$ for $m$-triplet. 
Furthermore, the SI symmetry requires that pair amplitudes belong to the even or odd parity representation, which is denoted by the index $g$ or $u$:
\begin{subequations}
\begin{align}
d^{\mu\nu}(\bk)&=(-)^P d^{\mu\nu}(-\bk)  & \mbox{for $\varGamma_g$ IRs}, \\
d^{\mu\nu}(\bk)&=(-)^{P+1} d^{\mu\nu}(-\bk) & \mbox{for $\varGamma_u$ IRs},
\end{align}
\end{subequations}
where $P=0$ for $\nu=0,z$ and is equal to the total parity of two orbitals $\ell$ and $\ell'$ for $\nu=x,y$. Therefore, the $m$-singlet/triplet pairing corresponds to the even/odd parity representation when the two orbitals have the same parity.

As a demonstration, let us mention a two-orbital system with $\varGamma_{6g}$ and $\varGamma_{7g}$ orbitals in $\rm D_{4h}$ point group. Both orbitals are twofold degenerate Kramers doublets. This two-orbital model has been studied as a minimal model of iron-based superconductors~\cite{zhou1,wan1}. The decomposition of direct products is given by $\varGamma_{6g}\otimes\varGamma_{6g}=\varGamma_{7g}\otimes\varGamma_{7g}=A_{1g}\oplus A_{2g}\oplus E_g$ and $\varGamma_{6g}\otimes\varGamma_{7g}=B_{1g}\oplus B_{2g}\oplus E_g$ (Table \ref{table1-D4}). Here, let us consider two examples of pairing states:
\begin{subequations}
\begin{align}
{\bm 0}~~ {\rm in}~~ \varGamma_{6g}\otimes\varGamma_{6g} ~~~(A_{1g}), \label{eq:A1g}\\
{\bm z}~~ {\rm in}~~ \varGamma_{6g}\otimes\varGamma_{7g} ~~~(B_{2g}). \label{eq:B2g}
\end{align}
\end{subequations}
These basis functions can be easily read from the third and the fourth column in Table \ref{table1-D4}. Next, we attach a function $\phi^{\varGamma}(\bk)$ in Table \ref{table1-D4} to the bases (\ref{eq:A1g}) and (\ref{eq:B2g}). 
For simplicity, we consider the following $\bk$ dependence:
\begin{subequations}
\begin{align}
\phi^{B_{1g}}(\bk)\,{\bm 0}~~ {\rm in}~~ \varGamma_{6g}\otimes\varGamma_{6g}~~~ (B_{1g}=B_{1g}\otimes A_{1g}),\label{eq:B1gA1g}\\
\phi^{A_{2g}}(\bk)\,{\bm z}~~ {\rm in}~~ \varGamma_{6g}\otimes\varGamma_{7g}~~~ (B_{1g}=A_{2g}\otimes B_{2g}).\label{eq:A2gB2g}
\end{align}
\end{subequations}
These two are both $B_{1g}$ IRs and we can find them in Table \ref{table2-D4}. However, they are not the final expression yet. Finally, we need to antisymmetrize Eqs.\,\eqref{eq:B1gA1g} and \eqref{eq:A2gB2g}. Equation \eqref{eq:B1gA1g} is already an antisymmetric expression, since $\phi^{B_{1g}}(\bk)$ is an even function and $\bm 0$ is antisymmetric (odd). As for Eq.\,\eqref{eq:A2gB2g}, it is necessary to antisymmetrize the orbital sector, $\varGamma_{6g}$ and $\varGamma_{7g}$. Since $\phi^{A_{2g}}(\bk)$ is even and $\bm z$ is symmetric (even), we should take an $o$-singlet $\tau^y$. Thus, we obtain the final form of the gap function with $B_{1g}$ $m$-singlet, $\phi^{A_{2g}}(\bk)\,\tau^y{\bm z}$. 
This is the outline to construct pair amplitudes with a specific IR in multi-orbital systems.

Before the end of this section, let us make some remarks on inter-orbital pairings in Tables \ref{table2-O} and \ref{table2-D6}.
One is that representations of some basis functions are {\it mixed}-parity and ambiguous. For example, $\phi^{A_1}(\bk)\times ({\bm z},i{\bm 0})$ belongs to $E_2$ representations of $\varGamma_7\otimes\varGamma_9$ in Table \ref{table2-D6}. Depending on $\phi^{A_1}=\phi^{A_{1g}}$ or $\phi^{A_{1u}}$, the basis functions are classified into two types of basis functions, 
\begin{subequations}
\begin{align}
&\phi^{A_{1g}}(\bk)\times (\tau^y{\bm z}, \tau^x{\bm 0})~~~~~ \mbox{($m$-singlet)}, \label{eq:kA1g} \\
&\phi^{A_{1u}}(\bk)\times (\tau^x{\bm z}, -\tau^y{\bm 0})~~~ \mbox{($m$-triplet)},\label{eq:kA1u}
\end{align}
\end{subequations}
after considering the fermion antisymmetry. 

Another is a special case in $\varGamma_{6(7)}\otimes\varGamma_8$ of O group in Table \ref{table2-O} as noted in Sec.\,\ref{sec:2b}. Since the pair can be $\varGamma_{6(7)}\otimes\varGamma_{8a}$ or $\varGamma_{6(7)}\otimes\varGamma_{8b}$, we need two kinds of $\tau$ matrices: one for $\varGamma_{6(7)}\otimes\varGamma_{8a}$ and the other for $\varGamma_{6(7)}\otimes\varGamma_{8b}$.

Tables \ref{table2-O}-\ref{table2-D6} are one of the main results in this paper. Even considering systems with two or more orbitals, the present results can be always applied by focusing on the $4\times4$ submatrix embedded in the entire space. Therefore, the basis functions in Tables \ref{table2-O}-\ref{table2-D6} are sufficient for any symmorphic systems. Although Tables \ref{table2-O}-\ref{table2-D6} seem to be rather complicated, they include important physical information about the pairing mechanism. This is because one can deduce what kinds (symmetry) of order parameters are realized when the system shows a characteristic fluctuation, since we have classified the superconducting order parameters in the orbital bases, which is easily related to the form of the characteristic interaction. In Sec.\,\ref{sec:3}, we will see this point by discussing several examples. 

\subsection{Band-based representations}\label{sec:2d}

So far, we have discussed the pair amplitudes and their basis functions in orbital-based representations. Here, let us examine the relation between the orbital-based and the band-based representations, since many observables strongly depend on the (band-based) energy gap on the Fermi surfaces.  

As usual, an intra-band Cooper pair amplitude can be defined by (the band index omitted),
\begin{align}
\tilde{F}_{\sigma\sigma'}(\bk)=\Big[\Big(\Phi(\bk)\sigma^0
+{\bm d}(\bk)\cdot {\boldsymbol{\sigma}} \Big)
i\sigma_y\Big]_{\sigma\sigma'}, \label{eq:dvec}
\end{align}
with {\it pseudo}-spin singlet amplitude $\Phi(\bk)$ and triplet ${\bm d}(\bk)$. 
Strictly, {\it pseudo}-spin $\sigma (\sigma')=\uparrow,\downarrow$ is the Kramers index for a given band. 
From Eqs.\,\eqref{eq:umat} and \eqref{eq:fdef}, one can obtain the relation between the band and the orbital-based pair amplitudes,
\begin{align} \label{eq:uuF}
\tilde{F}_{\sigma\sigma'}(\bk)=\sum_{\ell\alpha,\ell'\alpha'} 
u_{\ell\alpha,\sigma}^*(\bk)u_{\ell'\alpha',\sigma'}^*(-\bk)F_{\ell\alpha,\ell'\alpha'}(\bk). 
\end{align}

Before discussing the details, let us explain our phase convention. We use a convention that the degenerate pair for a given $\bk$ satisfies 
\begin{align}
(\varTheta I)c^\dag_{\ell\pm}(\bk)(\varTheta I)^{-1}=\mp c^\dag_{\ell\mp}(\bk),\label{TI}
\end{align}
under the time-reversal ($\varTheta$) and spatial inversion ($I$) operations. Using this convention, one obtains
\begin{subequations}\label{eq:uram}
\begin{align}
&u_{\ell +,\uparrow}(\bk)= (-1)^{P_\ell}u^*_{\ell -,\downarrow}(\bk),\\
&u_{\ell +,\downarrow}(\bk)= (-1)^{P_\ell+1}u^*_{\ell -,\uparrow}(\bk), 
\end{align}
\end{subequations}
where $P_\ell$ is the parity of the orbital $\ell$.
Furthermore, in centrosymmetric systems, one can take
\begin{align}
u_{\ell\alpha,\sigma}(\bk)=u_{\ell\alpha,\sigma}(-\bk) (-1)^{\bar{P}_\ell}, \label{Inv_u}
\end{align}
with $\bar{P}_\ell\equiv P_\ell+P_0$, where $P_0$ is the parity for a reference orbital $\ell_0$ of the band electron concerned (See the definition of $\ell_0$ below).

Although the sum of $\ell(\ell')$ in Eq.\,\eqref{eq:uuF} contains all of orbitals, it is sufficient to consider the case of two orbitals $\ell(\ell')=1,2$ in the discussion below. In Eq.\,\eqref{eq:Flm}, $F_{\ell\alpha,\ell'\alpha'}(\bk)$ is expressed by $d^{\mu\nu}_{\ell\ell'}(\bk)$, which is related to $\Phi(\bk)$ and  ${\bm d}(\bk)$ in the following way,
\begin{align}\label{eq:wdd}
\begin{pmatrix}
\Phi(\bk)\\
{\bm d}(\bk)\\
\end{pmatrix}
=(-1)^{\bar{P}_\ell}\sum_{s=\pm}\sum_{\nu=0,x,y,z}\!\!\!\!{\mathcal W}^s_\nu(\bk)
\begin{pmatrix}
d_{s}^{0\nu}(\bk)\\
\vec{d}_s^{\nu}(\bk)\\
\end{pmatrix},
\end{align}
with $[\vec{d}^{\nu}_s(\bk)]_{\mu}=d_s^{\mu\nu}(\bk)$, $d_\pm^{\mu\nu}=\tfrac{1}{2}(d^{\mu\nu}_{12}\pm d^{\mu\nu}_{21})$, and ${\mathcal W}^s_\nu(\bk)$ are transformation matrices defined below. 
When the two orbitals have the same parity $P_1=P_2$, due to the fermion antisymmetry, only ${\mathcal W}^+_{0,x,z}$ and ${\mathcal W}_y^-$ are nonvanishing, and the others are zero;
\begin{subequations}
\begin{align}
{\mathcal W}^+_{\nu}(\bk)=&
\begin{pmatrix}
w^0_{0\nu}&0 & 0 &0\\
\vec{0} & -\vec{w}_{x\nu} & \vec{w}_{y\nu} & -\vec{w}_{z\nu}\\
\end{pmatrix}, \label{eq:w1} \\
{\mathcal W}^-_y(\bk)=i&
\begin{pmatrix}
0 & -w^0_{xy}& w^0_{yy}& -w^0_{zy}\\
\vec{w}_{0y} & \vec{0} & \vec{0} & \vec{0} \\
\end{pmatrix}, \label{eq:w2}
\end{align}
\end{subequations}
where $\nu=0,x,$ and $z$. Here, $\vec{0}=(0,0,0)^T$ and 
\begin{subequations}
\begin{align}
& w^0_{\mu\nu}=(-1)^{P_\ell}(u \sigma^\mu\tau^\nu u^*), \\
& \vec{w}_{\mu\nu}=\Big[{\rm Re}(u \bar{\sigma}^\mu\tau^\nu u), {\rm Im}(u \bar{\sigma}^\mu\tau^\nu u), -w^0_{\mu\nu}\Big]^T,
\end{align}
\end{subequations}
with
\begin{align}
(u \sigma^\mu \tau^\nu u')\equiv \sum_{\alpha\alpha'}^\pm \sum_{\ell\ell'}^{1,2}
u_{\ell\alpha,\uparrow}(\bk)\sigma^\mu_{\alpha\alpha'}\tau^\nu_{\ell\ell'} u'_{\ell'\alpha',\uparrow}(\bk), \label{uumat}
\end{align}
and $\sigma^\mu \to \bar{\sigma}^\mu$.
Even when the two parities are different $P_1\ne P_2$, $\mathcal W^s_{\nu}$ can be easily obtained by multiplying $(-1)$ and replacing ${\mathcal W}^{\pm}_\nu\to {\mathcal W}^{\mp}_\nu$ in Eqs.\,\eqref{eq:w1} and \eqref{eq:w2}. Note also that in this case, ${\mathcal W}^{\pm}_\nu(\bk)=-{\mathcal W}^{\pm}_\nu(-\bk)$ holds from Eq.\,\eqref{Inv_u}.

Equation\,\eqref{eq:wdd} indicates that $\tilde{F}_{\sigma\sigma'}(\bk)$ is the product of $\mathcal W^s_\nu(\bk)$ and the orbital-based $F_{\ell\alpha,\ell\alpha'}(\bk)$. Thus, the $\bk$ dependence of ${\mathcal W}^\pm_{\nu}(\bk)$ can yield additional nodes in the band-based gap functions~\cite{Bishop}. We will discuss this aspect in Sec.\,\ref{sec:3}, but before that, we need to explain how to fix the phase ambiguity involved in $\mathcal{W}^\pm_\nu(\bk)$.

Generally, when the TR and SI symmetries are held, $\mathcal{W}^\pm_\nu(\bk)$ is accompanied by at least U(2) phase ambiguity for every band and at every $\bk$ point, due to the U(1) gauge and the Kramers degeneracy. In order to remove such ambiguity, a natural phase fixing procedure is necessary.
Here, we consider assigning an IR of the point group to each band $n$ in such a way that the IR corresponds to that of the dominant orbital $\ell_0$ for the band $n$. Indeed, the choices of the IRs are arbitrary, but the above choice is one of natural ways as explained below. This can be performed by the following procedure; for the dominant orbital component $\ell_0$ in the band $n$, $u_{\ell_0\pm,n\mp}(\bk)$ are set to zero and $u_{\ell_0\pm,n\pm}(\bk)$ to a real number, respectively (See Appendix \ref{app:3}). This way of the phase convention naturally connects generic situations to the orbital-diagonal limit, where there exist no hybridizations between different orbitals. With this, the band $n$ and the main orbital $\ell_0$ have the same symmetry without ambiguity. Therefore, Tables \ref{table2-O}-\ref{table2-D6} are still valid in the band-based Cooper pairs (See Appendix \ref{app:4}).

Using the phase-fixed bases, one can discuss the additional nodes through ${\mathcal W}^\pm_{\nu}(\bk)$. Information of the IR in the orbital-based Cooper pairs is encoded in ${\mathcal W}^\pm_{\nu}(\bk)$, and thus, ${\mathcal W}^\pm_{\nu}(\bk)$ can possess nodes if this belongs to an anisotropic IRs. Equation\,\eqref{eq:wdd} means that the $\bk$ dependence of the band-based pair amplitudes is determined by a product of ${\mathcal W}^\pm_{\nu}(\bk)$ and the orbital-based ones. This implies that even local orbital pairs can be transformed into anisotropic ones in the band representation, and also non-$A_{1g}$ inter-orbital pairs can lead to an anisotropic $A_{1g}$ band-based pairs in connection with non-$A_{1g}$ ${\mathcal W}^\pm_{\nu}(\bk)$. 
In the next section, we will discuss these mechanisms to realize anisotropic superconductivity in detail. 

\section{Examples}\label{sec:3}
In this section, we discuss (i) the pairing states emerging in close proximity to (anti-)ferroic quadrupole ordering, (ii) a mechanism of anisotropic $s$-wave ($A_{1g}$) pairing state, and (iii) anisotropic pairing states mediated by local fluctuations. 
The case (i) is a generalization of spin-fluctuation mechanism; $d$-wave pairing state~\cite{Miyake,Scalapino} next to antiferromagnetic phases, or $p$-wave to ferromagnetic phases. 
We will discuss these features unique to multi-orbital superconductors, which are our main results in this paper. Here, we focus on gap functions rather than the Cooper pair amplitudes, since the former can be more easily obtained in actual calculations.

\subsection{$\varGamma_8$ model in a cubic lattice}\label{sec:3a}
First, let us consider a model with non-Kramers doublet $\varGamma_{8u}$ on a simple-cubic lattice. It may be related to recently discovered superconductivity in Pr-based 1-2-20 compounds~\cite{Matsubayashi,Tsujimoto}. Local bases $|\varGamma_{8a,b};\pm\rangle$ are fourfold degenerate with the orbital $a$, $b$ and the Kramers degeneracy $\pm$. For simplicity, as a pairing interaction $H_{\rm int}$, we take the nearest-neighbor $E_g$-orbital (quadrupole) fluctuations,
\begin{align}
& H_{\rm int}=\frac{1}{N}\sum_{\bq}\sum_{i}v(\bq)\mathcal{M}_{E_g^i}(-\bq) \mathcal{M}_{E_g^i}(\bq) , \label{DensityDensity}\\
& \mathcal{M}_{E_g^i} (\bq)=\sum_{\bk}\sum_{12}\big[\hat{M}_{E_g^i}\big]_{12}c_{1}^{\dagger}(\bk)c_{2}(\bk+\bq),
\end{align}
where the sum of $1(2)$ symbolically represents the sum of the fourfold local bases $|\varGamma_{8a,b};\pm\rangle$, and the matrices of the multipole part $\hat{M}_{E_g^i}$ are defined by
\begin{align}
\hat{M}_{E_g^1}=\frac{\tau^z\sigma^0}{2},~~~~~ \hat{M}_{E_g^2}=\frac{\tau^x\sigma^0}{2}.
\end{align}
Thus, $[\tau^\nu\sigma^\mu]_{12}=\tau^\nu_{a_1a_2}\sigma^\mu_{\sigma_1\sigma_2}$ with $a_j=a$ or $b$ and $\sigma_j=\pm$. The momentum dependence of the pairing interaction is $v(\bq)=2v(c_x+c_y+c_z)$, where $v$ is a constant, $c_\mu=\cos q_\mu$ ($\mu=x,y,z$) and the lattice constant is set to unity. Note that the normalization condition ${\rm Tr}\big[\hat{M}_{E_g^i}\hat{M}_{E_g^j}^\dagger\big]=\delta_{ij}$ is satisfied, where Tr is taken for both the orbital and the Kramers indices.

Now, let us solve a superconducting gap equation within the mean-field theory (see Appendix \ref{app:0}). It is convenient to decouple Eq.\,\eqref{DensityDensity} into each Cooper channel. To this end, we rewrite $v(\bq)$ as follows,
\begin{align}
v(\bk-\bk')=v\sum_{\varGamma}\sum_i\phi^{\varGamma}_i(\bk)\phi^{\varGamma}_i(\bk'), \label{eq:3a-1}
\end{align}
where $\varGamma$ runs over $A_{1g}$, $E_g$, and $T_{2g}$ IRs, and $i$ is the label for different bases in $E_g$ and $T_{2g}$. The basis functions $\phi^{\varGamma}_i(\bk)$ are defined as follows,
\begin{subequations}
\begin{align}
& \phi^{A_{1g}} =\sqrt{\tfrac{2}{3}} \(c_x+c_y+c_z\), \\
& \phi^{E_g}_1 =\tfrac{1}{\sqrt{3}} \(2c_z-c_x-c_y\), \\
& \phi^{E_g}_2 =c_x-c_y, \\
& \phi^{T_{1u}}_\mu =\sqrt{2}s_\mu,\qquad(\mu=x,y,z)
\end{align}
\end{subequations}
with $s_\mu=\sin k_\mu$. These basis functions meet the orthonormality condition:
\begin{align}
\frac{1}{N}\sum_\bk\phi^{\varGamma}_i(\bk) \big(\phi^{\varGamma'}_j(\bk)\big)^*=\delta_{ij}\delta_{\varGamma\varGamma'}.
\end{align}
Then, we can decompose the pairing interaction into the zero-momentum Cooper channels, 
\begin{align}
&H_{\rm int}=-\frac{1}{2N}\sum_{\varGamma \alpha} \sum_i v^\varGamma_{\alpha} \; \Psi_{\varGamma i,\alpha}^\dagger \Psi_{\varGamma i,\alpha}, \label{eq:intchannel} \\
&\Psi_{\varGamma i,\alpha}^\dagger=\sum_{\bk}\sum_{12}\big[\hat{\varphi}^{\varGamma}_{\alpha,i}(\bk)\big]_{12} c^\dagger_1(\bk)c^\dagger_2(-\bk). 
\end{align}
Here, the form factor $\hat{\varphi}^{\varGamma}_{\alpha,i}(\bk)$, which will be calculated below and shown in Eqs.\,\eqref{eq:g8a} and \eqref{eq:g8b}, is regarded as a basis function of the Cooper channel labeled by $\varGamma$, $i$, and $\alpha$. The $\bk$ dependence of gap functions is determined by one or a linear combination of $\hat{\varphi}^{\varGamma}_{\alpha,i}(\bk)$ (Appendix \ref{app:0}).

For the decomposition into the Cooper channels, it is convenient to use the following identity
\begin{align}
2\sigma^0_{14}\sigma^0_{23}=\sum_\mu\bar{\sigma}^\mu_{12}\bar{\sigma}^{\mu*}_{43},
\end{align}
and similar ones for the orbital components. 
Signs arising from these decomposition are summarized in Table \ref{table3}, which is also useful to understand what kinds of Cooper channel are attractive. In the present case, we obtain the following decomposition,
\begin{align}
\begin{split}
&\sum_{i=z,x} [\tau^i\sigma^0]_{14}[\tau^i\sigma^0]_{23}= \\ 
&~~~~~\frac{1}{2}\sum_\mu\Big([\tau^0\bar{\sigma}^\mu]_{12}[\tau^0\bar{\sigma}^\mu]^*_{43}-[\tau^y\bar{\sigma}^\mu]_{12}[\tau^y\bar{\sigma}^\mu]^*_{43} \Big).\label{eq:decompCube}
\end{split}
\end{align}
This indicates that the pairing interaction is $v$ for $o$-singlet, and $-v$ for $o$-triplet. 

\begin{table}[b]
\caption{Signs $c^i_{\mu\nu}$ involved in the decomposition from particle-hole (ph) to the Cooper channels: $2\tau^\mu_{a_1a_4}\tau^\mu_{a_2a_3} =\sum_{\nu}c^1_{\mu\nu}\tau^\nu_{a_1a_2}\tau^{\nu *}_{a_4a_3}$ for the orbital sector, and $2\sigma^\mu_{\sigma_1\sigma_4}\sigma^\mu_{\sigma_2\sigma_3} =\sum_{\nu}c^2_{\mu\nu}\bar{\sigma}^\nu_{\sigma_1\sigma_2}\bar{\sigma}^{\nu *}_{\sigma_4\sigma_3}$ for the spin sector.}
\label{table3}
\begin{tabular}{>{\centering\arraybackslash}p{22mm}>{\centering\arraybackslash}p{14mm}>{\centering\arraybackslash}p{14mm}>{\centering\arraybackslash}p{14mm}>{\centering\arraybackslash}p{14mm}} \hline \hline
 ph-channels & \multicolumn{4}{c}{Cooper channels} \\ 
 & $\tau^0\tau^{0*}$ & $\tau^x\tau^{x*}$ & $\tau^y\tau^{y*}$ & $\tau^z\tau^{z*}$ \\ \hline
$2\tau^0\tau^0$ & $1$ & $1$ & $1$ & $1$ \\
$2\tau^x\tau^x$ & $1$ & $1$ & $-1$ & $-1$ \\
$2\tau^y\tau^y$ & $-1$ & $1$ & $-1$ & $1$ \\
$2\tau^z\tau^z$ & $1$ & $-1$ & $-1$ & $1$ \\
\hline \hline
& $\bar{\sigma}^0\bar{\sigma}^{0*}$ & $\bar{\sigma}^x\bar{\sigma}^{x*}$ & $\bar{\sigma}^y\bar{\sigma}^{y*}$ & $\bar{\sigma}^z\bar{\sigma}^{z*}$ \\ \hline
$2\sigma^0\sigma^0$ & $1$ & $1$ & $1$ & $1$ \\
$2\sigma^x\sigma^x$ & $-1$ & $-1$ & $1$ & $1$ \\
$2\sigma^y\sigma^y$ & $-1$ & $1$ & $-1$ & $1$ \\
$2\sigma^z\sigma^z$ & $-1$ & $1$ & $1$ & $-1$ \\
\hline
\end{tabular}
\end{table}

Now, let us illustrate a possible phase diagram. 
In the case of $v>0$ (antiferroic $E_g$ fluctuations), the $o$-singlet channels $\tau^y\bar{\sigma}^{\mu}$ in Eq.\,\eqref{eq:decompCube} is attractive. Thus, the gap functions for the following channels can be realized;
\begin{align}
\phi^{A_{1g}} \eta^{\nu}{\boldsymbol \mu},~ \phi^{E_g}_{1,2} \eta^{\nu}{\boldsymbol \mu},~   \phi^{T_{1u}}_\mu \tau^y{\bf 0}, \label{eq:g8a}
\end{align}
which belong to, respectively, $T_{2g}$, $T_{2g,1g}$, and $T_{2u}$ IRs in Table \ref{table2-O}.
Following the symmetrization procedure in Sec.\,\ref{sec:2c}, we find that ${\boldsymbol \mu}={\bm 0}$ components in Eq.\,\eqref{eq:g8a} are forbidden due to the fermion antisymmetry, since $\phi^{A_{1g}}(\bk)$ and $\phi^{E_g}(\bk)$ are even functions in $\bk$. 
Thus, it is natural that the superconducting states in close proximity to an antiferroic quadrupole ordered phase belong to three-dimensional representations. In this regards, it is very interesting to explore what kinds of superconducting state are realized in Pr-based 1-2-20 compounds under high pressures, where the quadrupole order is suppressed~\cite{Matsubayashi}.

Next, in the case of $v<0$ (ferroic quadrupole fluctuations), $o$-triplet channels $\tau^0\bar{\sigma}^{\mu}$ in Eq.\,\eqref{eq:decompCube} are favored. The gap functions in attractive channels are
\begin{align}
 \phi^{A_{1g}} \tau^0 {\bf 0},~ \phi^{E_g}_{1,2}\tau^0{\bf 0},~ \phi^{T_{1u}}_\mu \zeta^\nu {\boldsymbol \mu},\label{eq:g8b}
\end{align}
which belong to, respectively, $A_{1g}$, $E_g$, and $\{A_{1u}$, $E_u$, $T_{1u,2u}\}$ IRs. 
Again, the fermion antisymmetry requires ${\boldsymbol \mu}\ne {\bf 0}$ in Eq.\,\eqref{eq:g8b}. It should be noted that the intersite fluctuations can lead to an $A_{1g}$ pairing state.

\begin{figure}[t]
\centering
\includegraphics[width=7cm,clip]{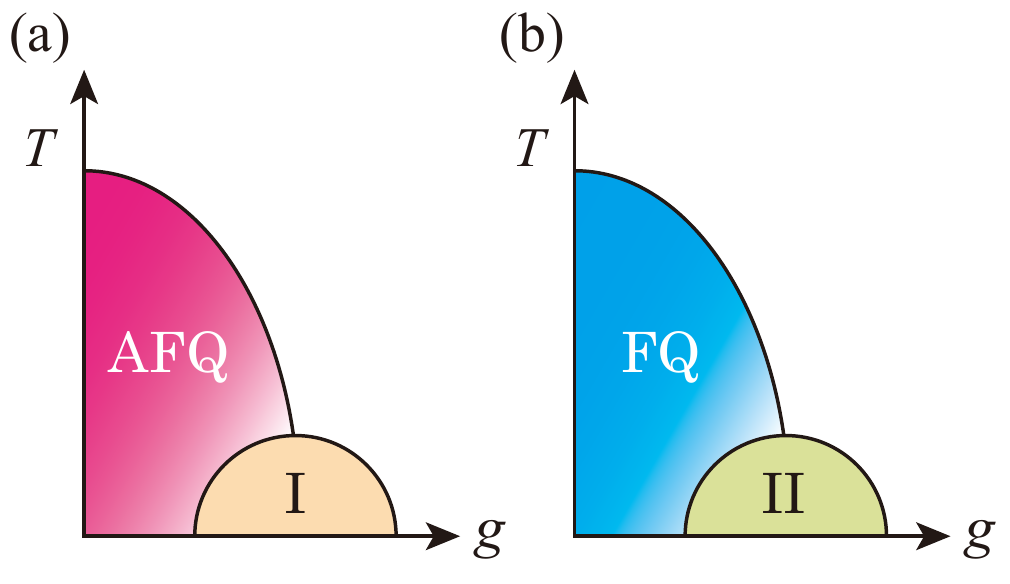}
\caption{(Color online) Schematic phase diagram near (a) antiferroic and (b) ferroic quadrupole ($E_g$) ordered phase as a function of temperature $T$ and a control parameter $g$, such as pressures. IRs of the obtained superconductivity belong to $T_{1g}, 2T_{2g}$, and $T_{2u}$ in the region \I\!\!, while $A_{1g}, A_{1u}, E_{g}, E_u, T_{1u}$, and $T_{2u}$ in the region \II.}
 \label{fig-1}
\end{figure}

Finally, let us illustrate schematic phase diagrams expected for antiferroic fluctuations in Fig.\,\ref{fig-1}(a) and for ferroic ones in Fig.\,\ref{fig-1}(b). The superconducting states in Fig.\,\ref{fig-1}(a) are expected to be three dimensional representations, while, in Fig.\,\ref{fig-1}(b), there are several candidates for the superconductivity within the present analysis. Fluctuations beyond the mean field approximation may favor some of the gap functions. Elaborated calculations are needed to clarify this. Note that the present results are based on a simple model, and the details depend on the electronic structures in actual materials.

It it often hard to observe quadrupole orderings experimentally. Several materials have been reported to exhibit quadrupole orders; CeB$_6$~\cite{shiina1,Effantin}, PrPb$_3$~\cite{Tayama,OnimaruPrPb3}, Pr-based 1-2-20 compounds~\cite{Onimaru} and so on~\cite{quadrupole}. As far as we know, among these systems, superconductivity is observed only in Pr-based 1-2-20 compounds~\cite{Matsubayashi,Tsujimoto}. Strictly speaking, as Pr-based 1-2-20 compounds are non-symmorphic systems, our theory is not directly applicable. However, the pressure-temperature phase diagram for PrV$_2$Al$_{20}$~\cite{MatsuUnpub} is similar to Fig.\,\ref{fig-1}(a). We can expect the emergence of unconventional three-dimensional superconductivity mentioned above.

\subsection{$\varGamma_{6u}$ and $\varGamma_{7u}$ model in a tetragonal lattice}\label{sec:3b}
The second example is a two-orbital model with $p_x / p_y$ orbitals in a two dimensional square lattice with $\rm D_{4h}$ symmetry. This corresponds to a model for BiS$_2$-layered superconductors, LaO$_{1-x}$F$_x$BiS$_2$~\cite{mizuguchi1}. Under $\rm D_{4h}$ symmetry, $p_x$ and $p_y$ orbitals are classified into $\varGamma_{6u}$ and $\varGamma_{7u}$; 
\begin{subequations}
\begin{align}
c^\dag_{\varGamma_{6u},\pm}=\frac{1}{\sqrt{2}}\left(ic^\dag_{p_x,\mp}\mp c^\dag_{p_y,\mp}\right), \\
c^\dag_{\varGamma_{7u},\pm}=\frac{1}{\sqrt{2}}\left(ic^\dag_{p_x,\mp}\pm c^\dag_{p_y,\mp}\right).
\end{align}
\end{subequations}
Here, $c^\dagger_{p_{x,y},\sigma}$ is the creation operator for the $p_{x,y}$ orbital with the {\it pure}-spin $\sigma=\pm$, while $c^\dagger_{\varGamma_{6u,7u},\alpha}$ is that for the $\varGamma_{6u,7u}$ orbital with the Kramers degrees of freedom $\pm$. In terms of $c^\dagger_{\varGamma_{6u,7u},\alpha}$, we define the non-interacting Hamiltonian by
\begin{align}
H_0=\sum_\bk\sum_{12} \big[\hat{h}(\bk)\big]_{12} c^\dag_{1}(\bk) c_{2}(\bk),\label{eq:bis2}
\end{align}
with
\begin{align}
\hat{h}(\bk)=\Big(h_0(\bk)\tau^0+\Delta\tau^z+h_x(\bk)\tau^x \Big)\sigma^0 +h_y(\bk)\tau^y\sigma^z.
\label{eq:3b-1}
\end{align}
Following Ref.\,\cite{usui1}, we set,
\begin{subequations}
\begin{align}
& h_0(\bk)=t_1(c_x+c_y)+t_2c_xc_y+t_3(c'_xc_y+c_x c'_y)-\mu, \\
& h_x(\bk)=t_4(c_x-c_y),  \\
& h_y(\bk)=[t_5+t_6(c_x+c_y)]s_xs_y, 
\end{align}
\end{subequations}
where $c'_{x,y}=\cos 2k_{x,y}$ and $(t_1,t_2,t_3,t_4,t_5,t_6,\mu)=(-0.334, 1.948, 0.166, -0.214,-1.572, -0.220, -1.40)$ in the unit of eV. 
The additional $\Delta$ term in Eq.\,\eqref{eq:3b-1} simply comes from the atomic SOC for the Bi $p$-electrons, and we set $\Delta=-0.15$. 
Note that the model \eqref{eq:bis2} holds $\rm D_{4h}$ symmetry, although the actual material LaO$_{1-x}$F$_x$BiS$_2$ belongs to non-symmorphic space group. 
Hereafter, by using the model \eqref{eq:bis2}, we discuss unconventional superconductivity due to two kinds of pairing mechanisms: (A) an inter-site orbital density wave fluctuations~\cite{lee2,athauda1}, and (B) a local fluctuation, {\it e.g.},  driven by electron-phonon interactions. 

First, let us consider fourfold-symmetry breaking orbital fluctuations. For simplicity, we consider $B_{1g}$ and $B_{2g}$ type orbital fluctuations, which are respectively described by $\hat{M}_{B_{1g}}=\tau^x\sigma^0/2$ and $\hat{M}_{B_{2g}}=\tau^y\sigma^z/2$ in $\varGamma_{6u}\otimes\varGamma_{7u}$ space. The corresponding pairing interaction is given by
\begin{align}
H_{\rm int}=\frac{1}{N}\sum_{\bq}\sum_{\varGamma=B_{1g},B_{2g}} v^\varGamma(\bq) \mathcal{M}_\varGamma(-\bq) \mathcal{M}_\varGamma(\bq), \label{eq:3b-2} 
\end{align}
with $v^\varGamma(\bq)=2v^\varGamma(c_x+c_y)$.
For $\bq=\bk-\bk'$, $v^\varGamma(\bk-\bk')$ can be decomposed into $A_{1g}$, $B_{1g}$, and $E_u$ IRs:
\begin{subequations}
\begin{align}
 \phi^{A_{1g}}&=c_x+c_y, \\
 \phi^{B_{1g}}&=c_x-c_y, \\
 (\phi^{E_u}_1,\phi^{E_u}_2)&=\sqrt{2}(s_x,s_y).
\end{align}
\end{subequations}
Thus, Eq.\,\eqref{eq:3b-2} simply reads
\begin{align}
\begin{split}
H_{\rm int}=-\frac{1}{4N}\sum_{\mu\nu}v^{\mu\nu}\sum_{1234}\big[\tau^\nu\bar{\sigma}^\mu\big]_{12}\big[\tau^\nu\bar{\sigma}^\mu\big]^*_{43}~~~~~~~~~~~~ \\
\times \sum_{\bk\bk'}\sum_{\varGamma i}\phi^\varGamma_i(\bk)\phi^\varGamma_i(\bk')c^\dag_1(\bk) c^\dag_2(-\bk) c_3(-\bk') c_4(\bk'), \label{eq:3b-3}
\end{split}
\end{align}
with $\varGamma=A_{1g}, B_{1g}$, or $E_u$. Here, $v^{\mu\nu}$ are given as follows,
\begin{subequations}
\begin{align}
4v^\I & =-(v^{B_{1g}}+v^{B_{2g}}), \\
4v^\II & =-(v^{B_{1g}}-v^{B_{2g}}), \\
4v^\III & =~~(v^{B_{1g}}-v^{B_{2g}}), \\
4v^\IV & =~~(v^{B_{1g}}+v^{B_{2g}}),
\end{align}
\end{subequations}
where the indices $\I\sim \IV$ indicate the following sets of $(\mu,\nu)$:
\begin{subequations}
\begin{align}
\I&: (0,0),~ (z,0),~ (x,x),~ (y,x), \label{eq:I}\\ 
\II&: (x,0),~ (y,0),~ (0,x),~ (z,x),\\ 
\III&: (0,y),~ (z,y),~ (x,z),~ (y,z),\\
\IV&: (x,y),~ (y,y),~ (0,z),~ (z,z).\label{eq:IV}
\end{align}
\end{subequations}
From the same analysis as in Sec.\,\ref{sec:3a}, for example, $\phi(\bk)\tau^0{\bm 0}$ is favored for the ferroic $B_{1g}/B_{2g}$ fluctuations, while $\phi(\bk)\tau^0{\bm x}$ for the ferroic $B_{1g}$ and the antiferroic $B_{2g}$ fluctuations, and so on. 
When we focus on even-parity pairing states, the gap functions favored by the present interactions are listed as follows:
\begin{subequations}
\begin{align}
\I&: ~~ \hat{\varphi}^{A_{1g}}_1=\phi^{A_{1g}}\tau^0{\bm 0},~~ \hat{\varphi}^{B_{1g}}_1=\phi^{B_{1g}}\tau^0{\bm 0},\\ 
\II&: ~~ \hat{\varphi}^{A_{1g}}_2=\phi^{B_{1g}}\tau^x{\bm 0},~~ \hat{\varphi}^{B_{1g}}_2=\phi^{A_{1g}}\tau^x{\bm 0},\label{eq:phib}\\
\III&: ~~ \hat{\varphi}^{A_{2g}}=\phi^{B_{1g}}\tau^y{\bm z},~~ \hat{\varphi}^{B_{2g}}=\phi^{A_{1g}}\tau^y{\bm z},\\
\IV&: ~~ \hat{\varphi}^{A_{1g}}_3=\phi^{A_{1g}}\tau^z{\bm 0},~~ \hat{\varphi}^{B_{1g}}_3=\phi^{B_{1g}}\tau^z{\bm 0}, \nonumber \\
&~~\, (\hat{\varphi}^{E_g}_{1,1},\hat{\varphi}^{E_g}_{1,2})=\phi^{A_{1g}}\tau^y(-{\bm x},{\bm y}), \nonumber \\
&~~\, (\hat{\varphi}^{E_g}_{2,1},\hat{\varphi}^{E_g}_{2,2})=\phi^{B_{1g}}\tau^y({\bm x},{\bm y}).
\end{align}
\end{subequations}
These orbital-based gap functions $\hat{\varphi}^{\varGamma}_i$ are transformed into the band-based ones $\tilde{\varphi}^{\varGamma}_i$ via unitary transformations as discussed in Sec.\,\ref{sec:2d}. It should be noted that the band-based $\tilde{\varphi}^{\varGamma}_i$ is crucially important in low-energy excitations observed experimentally. In what follows, let us elucidate the nodal structure of $\tilde{\varphi}^{\varGamma}_i$. 

\begin{figure}[t]
\centering
\includegraphics[width=8.5cm]{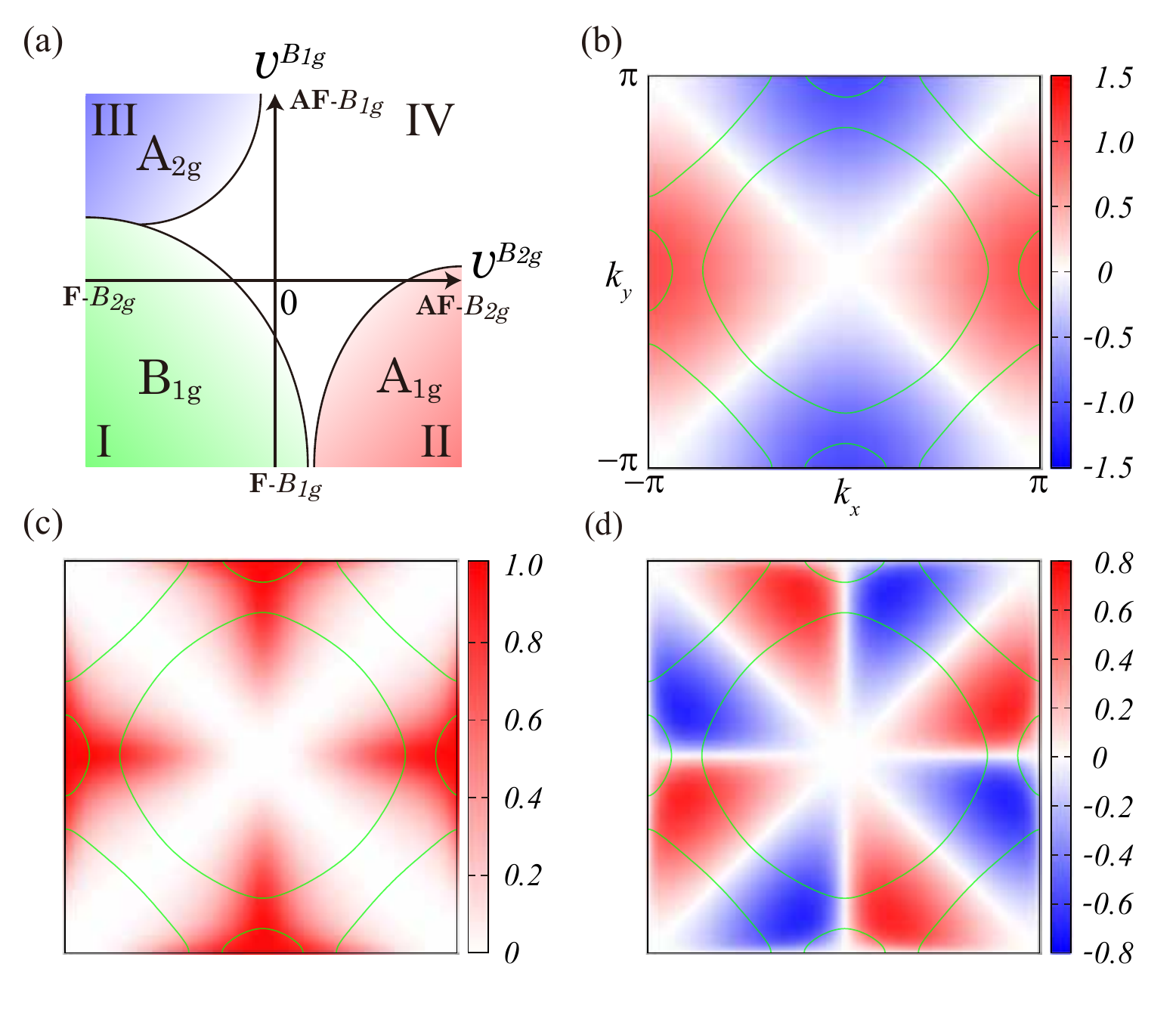}
\caption{(Color online) (a) Schematic phase diagram of the simple two-orbital model for BiS$_2$ layered superconductors. $B_{1g}$ and $B_{2g}$ type ferroic(F)/antiferroic(AF) orbital fluctuations have been considered. A typical band-based gap structure $\tilde{\varphi}^\varGamma(\bk)$ is illustrated in (b) $\varGamma=B_{1g}$, (c) $A_{1g}$, and (d) $A_{2g}$ states. The (green) solid lines indicate the Fermi surface in this model. }
\label{fig:3-2}
\end{figure}

For the case I, the nodal structures of $\tilde{\varphi}$'s solely come from those in $\hat{\varphi}^{A_{1g}}_1$ or $\hat{\varphi}^{B_{1g}}_1$, since $\tau^0{\bm 0}$ is $A_{1g}$. In contrast, in the case \II, due to a unique property of multi-orbital systems, both $\tilde{\varphi}^{A_{1g}}_2$ and $\tilde{\varphi}^{B_{1g}}_2$ possess nontrivial nodal structure along $k_x\pm k_y=0$ lines. In the orbital-based $\hat{\varphi}^{B_{1g}}_2$, since the $\bk$ dependence of $\phi^{A_{1g}}$ belongs to $A_{1g}$, the nodal structures of $\tilde{\varphi}^{B_{1g}}_2$ come from the unitary matrix through Eq.\,\eqref{eq:wdd}. Indeed, $\tau^x{\bm 0}$ is $B_{1g}$ IR in Table \ref{table1-D4}. The elements of the unitary matrix, which transform into the band mainly composed of $\varGamma_{6u}$ orbital, are given by
\begin{align}
\(u_{\varGamma_{6u}+,\uparrow},~ u_{\varGamma_{6u}-,\uparrow}\) & \sim \(1, 0\),  \nonumber \\
\(u_{\varGamma_{7u}+,\uparrow},~u_{\varGamma_{7u}-,\uparrow}\) & \sim \(e^{2i\theta_{\bk}}, e^{-i\theta_{\bk}}\), \nonumber
\end{align}
with $\theta_{\bk}$ being the angle in the $k_x$-$k_y$ plane. Then,
\begin{align}\label{eq:3b-4}
[{\mathcal W}^+_x(\bk)]_{11} \sim \cos 2\theta_{\bk}\sim k_x^2-k_y^2,
\end{align}
which has, indeed, $B_{1g}$ symmetry (Appendix \ref{app:4}). As for the gap function with $A_{1g}$ symmetry, it is commonly considered that it does not have symmetry-protected nodes. However, for $\hat{\varphi}^{A_{1g}}_2$ in Eq.\,\eqref{eq:phib}, since both $\phi^{B_{1g}}$ and Eq.\,\eqref{eq:3b-4} have line nodes along $k_x\pm k_y=0$, $\tilde{\varphi}^{A_{1g}}_2$ possesses $B_{1g}$-like gap nodes [Fig.\,\ref{fig:3-2}(c)]. Although these nodes are not symmetry protected, one can expect that a specific fluctuation leads to such accidental nodes in $A_{1g}$ gap functions.

Figure\,\ref{fig:3-2}(a) depicts the schematic phase diagram for the even parity sector obtained by numerical calculations. The corresponding nodal structures are summarized in Figs.\,\ref{fig:3-2}(b)-(d). The region around \IV is regarded as a normal state, because the corresponding $T_c$ is very low due to the fact that the attractive pairs are in inter-band pairing rather than intra-band pairing. Figure\,\ref{fig:3-2}(c) clearly shows that the $A_{1g}$ gap function is strongly anisotropic as discussed above.
It should be emphasized that this orbital-driven anisotropic $A_{1g}$ gap is not specific to the present model, but can commonly appear in any multi-orbital superconductors. 
This mechanism may provide a clue to understanding gap anisotropies in, {\it e.g.,} CeRu$_2$~\cite{Kittaka} and PrOs$_4$Sb$_{12}$~\cite{Izawa,Matsuda}.
Furthermore, the appearance of the $A_{2g}$ gap structure can be also regarded as a characteristic
property of multi-orbital systems, because if the $\bk$ dependence of the gap function comes only from $\phi^{A_{2g}}(\bk)$, $\phi^{A_{2g}}(\bk)$ must take the form of $\phi^{A_{2g}}(\bk)\sim\sin k_x\sin k_y(\cos k_x -\cos k_y)$. To realize such gap function in a single orbital system, there need much longer-range interactions than in the present nearest-neighbor model.

\subsection{Nodal gap derived from local fluctuations}\label{sec:3c}
\begin{figure}[t]
\centering
\includegraphics[width=8.5cm]{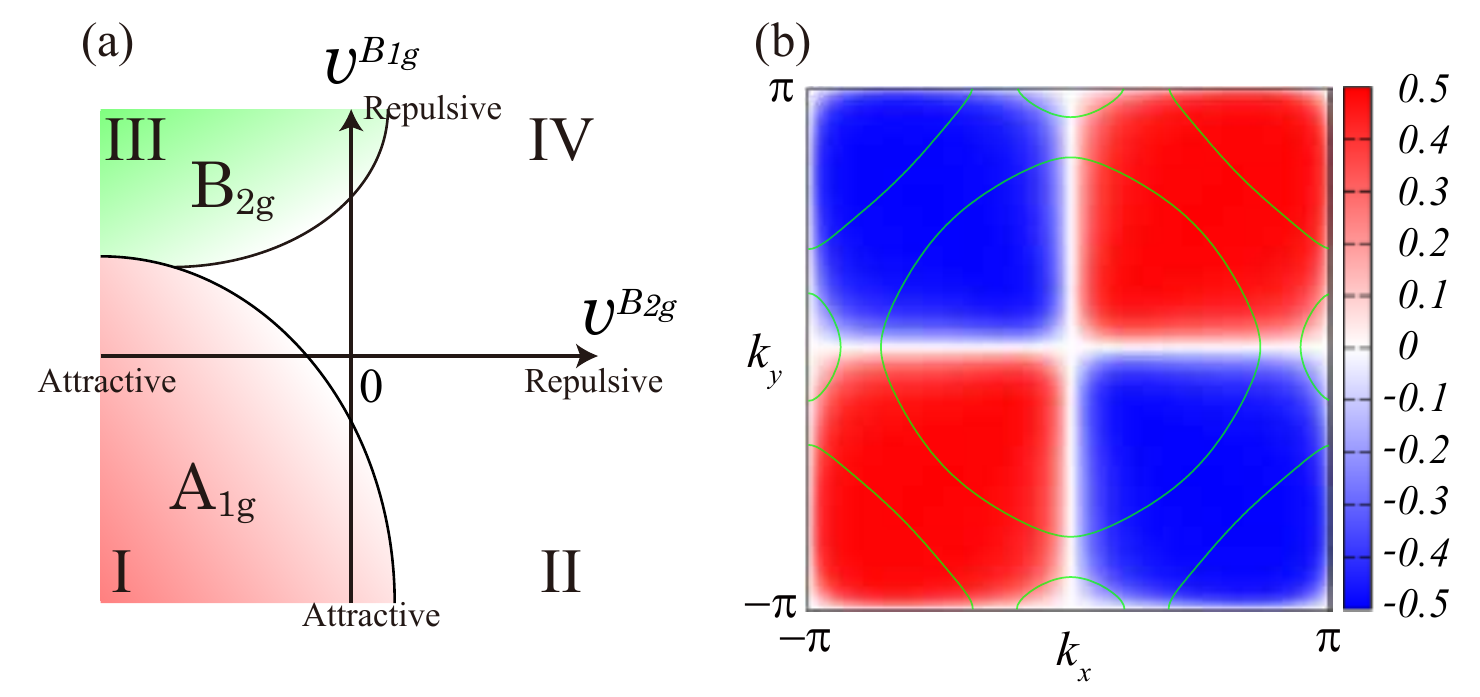}
\caption{(Color online) (a) Schematic phase diagram for local (on-site) $B_{1g}$ and $B_{2g}$ fluctuations in the same model as in Fig.\,\ref{fig:3-2}. (b) The momentum dependence of the band-based gap function for $B_{2g}$ state appearing around the region \III.}
\label{fig:3-3}
\end{figure}

Next, we focus on local fluctuations with no $\bk$ dependence. First, we show that only the local fluctuations can induce anisotropic and nodal superconductivity~\cite{Bishop} in sharp contrast to a naive expectation. As in Sec.\,\ref{sec:3b}, we consider $B_{1g}$ and $B_{2g}$ fluctuations, setting a constant $v^\varGamma(\bq)=v^\varGamma$ in Eq.\,\eqref{eq:3b-2}. 
In this case, the basis functions in the orbital basis are also independent of $\bk$. Therefore, the possible gap functions in attractive channels are:
\begin{subequations}
\begin{align}
\I&:~~ \hat{\varphi}^{A_{1g}}_{1}=\tau^0{\bm 0},\\ 
\II&:~~ \hat{\varphi}^{B_{1g}}=\tau^x{\bm 0},\\
\III&:~~ \hat{\varphi}^{B_{2g}}=\tau^y{\bm z},\\
\IV&:~~ \hat{\varphi}^{A_{1g}}_{2}=\tau^z{\bm 0},~ (\hat{\varphi}^{E_g}_1,\hat{\varphi}^{E_g}_2)=\tau^y(-{\bm x},{\bm y}),
\end{align}
\end{subequations}
where $\I \sim \IV$ represent the regions specified in Eqs.\,\eqref{eq:I}-\eqref{eq:IV}. 
Note that any odd parity $\hat{\varphi}^{\varGamma_u}$ is not allowed in stark contrast to the cases in Sec.\,\ref{sec:3b}. As typical examples, we focus on the $\tilde{\varphi}^{B_{1g}}$ and $\tilde{\varphi}^{B_{2g}}$. As mentioned in Sec.\,\ref{sec:3b}, $\tilde{\varphi}^{B_{1g}}$ and $\tilde{\varphi}^{B_{2g}}$ in the band representation must have nodes. The $\bk$ dependence of $\tilde{\varphi}^{B_{1g}} (\tilde{\varphi}^{B_{2g}})$ come from $u(\bk)$ and shows line nodes along $k_x\pm k_y=0$ ($k_xk_y=0$).

Figure \ref{fig:3-3}(a) depicts the $v^{B_{2g}}$-$v^{B_1g}$ phase diagram. We find that, due to only local fluctuations, anisotropic $B_{2g}$ gap structure can emerge in the region around \III\!\!\!. The obtained $B_{2g}$ nodal structure is illustrated in Fig.\,\ref{fig:3-3}(b). In particular, for large repulsion of $v^{B_{1g}}$, the nodal superconductivity with $B_{1g}$ symmetry can be induced by only repulsive local interactions. This can be understood via Table \ref{table3}; the repulsion in $B_{1g}$-channel leads to the attractive interaction in $B_{2g}$ channel. 
Thus, in multi-orbital systems, anisotropic gap structure can be also realized in the BCS approximation of purely local (on-site) interactions. This implies that an electron-phonon (e-ph) can lead to unconventional superconductivity.
In what follows, let us elucidate local fluctuations arising from e-ph couplings. 

In general, a specific phonon mode couples to electronic multipoles with the same IR. Local nonmagnetic multipoles in the present two-orbital model with $\varGamma_{6u}$ and $\varGamma_{7u}$ are written as
\begin{align}
\mathcal{M}_\varGamma(\br)=\sum_{12}\big[\hat{M}_\varGamma\big]_{12}c_1^\dag(\br)c_2(\br),
\end{align} 
with $\varGamma=A_{1g},B_{1g}$, or $B_{2g}$, and $\hat{M}_\varGamma$'s are given by
\begin{align}
\hat{M}_{A_{1g}}=\frac{\tau^0\sigma^0}{2},~~
\hat{M}_{B_{1g}}=\frac{\tau^x\sigma^0}{2},~~
\hat{M}_{B_{2g}}=\frac{\tau^y\sigma^z}{2}.
\end{align}
Integrating out the phonon degrees of freedom, we obtain an effective interaction, 
\begin{align}
H_{\rm int}=-\sum_\varGamma\frac{g_\varGamma^2}{\Omega^\varGamma} 
\sum_{\br}
 \mathcal{M}_\varGamma(\br)\mathcal{M}_\varGamma(\br),\label{eq:3c-1}
\end{align}
where $g_{\varGamma}$ is the local e-ph couplings and $\Omega^\varGamma$ is the local phonon frequency for $\varGamma=A_{1g}$, $B_{1g}$, and $B_{2g}$ mode. 
Following the procedure in Secs. \ref{sec:3b} and \ref{sec:3c}, Eq.\,\eqref{eq:3c-1} can be decomposed in the same way as in Eq.\,\eqref{eq:3b-3} with appropriate $v^{\mu\nu}$.
Using Table \ref{table3}, we obtain {\it e.g.,} $4v^{0x}=v^{A_{1g}}+v^{B_{1g}}-v^{B_{2g}}$, and so on. Note that such interactions $v^\varGamma=g_\varGamma^2/\Omega^\varGamma$ are always positive, different from the electron-electron interactions.
Therefore, since $A_{1g}$ pairing channel $\tau^0{\bm 0}$ is always attractive in all the phonon modes, namely, $4v^{00}=\sum_{\varGamma}v^\varGamma$, we re-realize that a fully-gapped $A_{1g}$ state is the most favorable.

One possibility of e-ph mediated anisotropic superconductivity arises when the Hund's coupling and the pair hopping term of on-site Coulomb repulsions are taken into account. For example, local interactions are $4v^{00}=U+J$ for $\tau^0{\bf 0}$ pairing, and $4v^{0x}=4v^{zy}=U-J$ for $\tau^x{\bm 0}$ and $\tau^y{\bm z}$, with the intra-orbital repulsion $U$ and the Hund's coupling $J$. Thus, the presence of the on-site Coulomb repulsions works against the isotropic pairing state as is well known. Another possibility is the $\bk$-dependent interaction via the e-ph coupling, but here we do not go into detail. 
Instead, let us focus on the fact that {\it e.g.}, $\tau^0\tau^0 \times \sigma^z\sigma^z = -1$ in Table \ref{table3}, which indicates that TR symmetry-breaking mode can suppress $A_{1g}$ pairing states. It implies that the e-ph interaction may lead to anisotropic pairing states in a magnetically-ordered state. These mechanisms for e-ph driven anisotropic superconductivity in combinations with other degrees of freedom are fascinating issues and we leave the detailed analysis in our future works.

\section{Conclusion}\label{sec:4}
We have constructed a complete table of irreducible representations of superconducting gap functions in symmorphic multi-orbital systems.  Classification in the orbital-based pairing (gap) functions offers novel entries in the classification tables. The Cooper pairs in multi-orbital systems can be regarded as ones with multipole degrees of freedom, and we have called it ``multipole superconductivity''.
From this viewpoint, we find that unconventional (anisotropic) superconductivity can be realized not only by the momentum dependence of the pairing interactions, but also by the orbital degrees of freedom. 

One of the nontrivial results appears in the system composed of $\varGamma_9 \otimes \varGamma_9$ orbitals in $D_6$ group. The transformation properties of the Cooper pairs are not explained by those for the {\it pure}-spin 1/2 in the conventional classification. This is an important consequence of orbital degrees of freedom. 

We have also clarified how the superconducting gap nodes appear in multi-orbital systems. We have explained the relation between the gap functions in the orbital bases and those in the band ones. The momentum dependence of the band-based gap functions depends on that of the orbital-based ones and the unitary matrix transforming the two bases. The latter depends on the IR for the corresponding Kramers degrees of freedom in the orbital bases, which include both the {\it pure}-spin and the orbital angular momentum and are generally not only the {\it pure}-spin 1/2.

On the basis of the present group theoretical analysis, we have discussed a cubic $\varGamma_{8u}$ model and tetragonal $\varGamma_{6u}+\varGamma_{7u}$ models. In the former model, superconductivity with anisotropic three-dimensional representations emerges in the vicinity of an antiferroic quadrupole ordered phase. 
In the latter, we have discussed the formation of anisotropic gap functions including anisotropic $s$-wave ($A_{1g}$) type functions induced by various orbital fluctuations. 
We have also proposed nodal/anisotropic superconductivity mediated by local fluctuations, which can be realized only in the multi-orbital systems.
Our findings imply that fluctuations arising from e-ph couplings also may induce anisotropic superconductivity with the help of the TR symmetry-breaking and the local Coulomb interactions, although the conventional local e-ph interactions favor isotropic $s$-wave pairing. We hope that the present study provides a renewed interest in multi-orbital systems and encourages experimental research for new superconducting materials.

\section*{Acknowledgments}
We acknowledge Y. Yanase for valuable discussions, and K. Izawa for providing his recent data and helpful discussions. This work was partly supported by JSPS KAKENHI Grant Nos. 15H05745, 15H02014, 15J01476, 16H01079, 16H01081, 16H04021.

\begin{appendix}

\section{General consideration of classification}
In the main text, we have classified superconducting gap functions according to IRs of a given point group P. Here, we show that superconducting order parameters can be characterized by IRs of P in both symmorphic and non-symmorphic space group G. Moreover, in symmorphic systems, the form of gap functions can be determined by considering only the spin-orbital coupled degrees of freedom.

\subsection{Classification under space group G}\label{app:0}
Let us consider a BCS type model Hamiltonian, $H=H_0+H_{\rm int}$, under a space group G,
\begin{align}
& H_0 =\sum_{\bk}\sum_{12}\big[\hat{h}(\bk)]_{12}c^\dag_1(\bk)c_2(\bk), \label{eq:hami} \\
& H_{\rm int}=-\frac{1}{2N}\sum_{\bk\bk'}\sum_{1234}v_{14,32}(\bk-\bk') \nonumber 
\\ & \hspace{2.5cm}\times c^\dag_1(\bk)c^\dag_2(-\bk)c_3(-\bk')c_4(\bk'), \label{eq:int}
\end{align}
where $\hat{h}(\bk)$ is a Hermitian matrix describing the band structure, and the subscripts ($1\sim4$) symbolically represent the orbital, the spin, and the atomic site degrees of freedom. 
In this Hamiltonian, $H_0$ and $H_{\rm int}$ should respectively be invariant under any operation $g$ in the space group G. That is to say, $[H_0,g]=0$ and $[H_{\rm int},g]=0$.
The space group element $g$ is denoted as $g=\{p|\ba\}$ in Seitz notation, where $p$ is an operation of the point group P associated with G, and $\ba$ is a translation. 
From $[H_0,g]=0$, we obtain, 
\begin{align}
\hat{U}(g;\bk)\hat{h}(\bk)\hat{U}^\dag(g;\bk)=\hat{h}(p\bk), \label{eq:uhu}
\end{align}
using the following relation,
\begin{align}
g\,c^\dag_{1}(\bk)\,g^{-1} = \sum_2 c^\dag_2(p\bk)[\hat{U}(g;\bk)]_{21},\label{eq:csym}
\end{align}
where the matrix $\hat{U}(g;\bk)$ describes the transformation property of $c_1^\dag(\bk)$, which generally depends on $\bk$. 
As for $H_{\rm int}$, one can expand $v_{14,32}(\bk-\bk')$ into the following form, 
\begin{align}
v_{14,32}(\bk-\bk')
=\sum_\varGamma\sum_i v^\varGamma
\big[\hat{\varphi}^\varGamma_i(\bk)\big]_{12}
\big[\hat{\varphi}^\varGamma_i(\bk')\big]^*_{43}. \label{eq:vpp}
\end{align}
Here, the sum of $\varGamma$ contains non-equivalent IRs of P and the label $i$ denotes degenerate bases in the same $\varGamma$.
$v^\varGamma$ can be regarded as a pairing interaction in $\varGamma$ IR channel, which is a real number due to the Hermitian of $H_{\rm int}$.
The matrix $\hat{\varphi}^\varGamma_i(\bk)$ is the $i$th basis function for the $\varGamma$ IR of P, which transforms according to, 
\begin{align}
\hat{U}(g;\bk)\hat{\varphi}^\varGamma_i(\bk)\hat{U}^T(g;-\bk)=\sum_j \hat{\varphi}^\varGamma_j(p\bk)\mathcal{D}_{ji}^{(\varGamma)}(p), \label{eq:phisym}
\end{align}
where $\mathcal{D}_{ji}^{(\varGamma)}(p)$ is the representation matrix of $\varGamma$ IR. Equation\,\eqref{eq:phisym} can be obtained from a requirement that 
\begin{align}
\Psi_{\varGamma i}^\dagger=\sum_\bk\sum_{12}
\big[\hat{\varphi}^\varGamma_i(\bk)\big]_{12}c^\dag _1(\bk)c^\dag_2(-\bk),\label{eq:g1}
\end{align}
satisfies the following transformation properties,
\begin{align}
g\,\Psi_{\varGamma i}^\dagger\,g^{-1}=\sum_{j}\Psi_{\varGamma j}^\dagger \mathcal{D}_{ji}^{(\varGamma)}(p).
\end{align}
Thus, $H_{\rm int}$ is written as follows,
\begin{align}
H_{\rm int}=-\frac{1}{2N}\sum_\varGamma\sum_i v^{\varGamma} \Psi_{\varGamma i}^\dagger \Psi_{\varGamma i}.
\end{align}
This clearly shows that $H_{\rm int}$ is certainly invariant under any operation $g$. 

Now, let us confirm the requirements of basis functions:
\begin{subequations}
\begin{align}
 \hat{\varphi}^\varGamma_i(\bk)=-\big(\hat{\varphi}^\varGamma_i(-\bk)\big)^T, \label{eq:breq1}\\
 \frac{1}{N}\sum_\bk {\rm Tr}\big[\hat{\varphi}^\varGamma_i(\bk)\hat{\varphi}^{\varGamma'\dag}_j(\bk)\big]=\delta_{ij}\delta_{\varGamma\varGamma'}.\label{eq:breq2}
\end{align}
\end{subequations}
The first equation \eqref{eq:breq1} is evident from Eq.\,\eqref{eq:vpp}, while the second one \eqref{eq:breq2} can be derived by using the grand orthogonal theorem among IRs;
\begin{align*}
& \sum_\bk{\rm Tr}
\big[\hat{\varphi}_i^\varGamma(\bk)\hat{\varphi}_j^{\varGamma'\dag}(\bk)\big]\\
&~~=\frac{1}{m}\sum_\bk\sum_p{\rm Tr}
\big[\hat{\varphi}_i^\varGamma(p\bk)\hat{\varphi}_j^{\varGamma'\dag}(p\bk)\big]\\
&~~=\frac{1}{m}\sum_\bk\sum_{i'j'}{\rm Tr}
\big[\hat{\varphi}_{i'}^\varGamma(\bk)\hat{\varphi}_{j'}^{\varGamma'\dag}(\bk)\big]
\sum_p \big(\mathcal{D}_{ii'}^{(\varGamma)}(p)\big)^* \mathcal{D}_{jj'}^{(\varGamma')}(p)\\
&~~=\frac{1}{d_\varGamma}\delta_{ij}\delta_{\varGamma\varGamma'}
\sum_\bk\sum_{i}{\rm Tr}
\big[\hat{\varphi}_{i}^\varGamma(\bk)\hat{\varphi}_{i}^{\varGamma\dag}(\bk)\big],
\end{align*}
where $m$ is the order of P, and $d_\varGamma$ the dimension of $\varGamma$. Thus, with the appropriate normalization, we can choose $\hat{\varphi}^\varGamma_i(\bk)$ to satisfy Eqs.\,\eqref{eq:breq1} and \eqref{eq:breq2}.

Next, we apply the mean-field theory to Eq.\,\eqref{eq:int}, and introduce the superconducting order parameter,
\begin{align}
\big[\hat{\Delta}(\bk)\big]_{12}
&=\frac{1}{N}\sum_{\bk'}\sum_{34} v_{14,32}(\bk-\bk')\langle c_4(\bk')c_3(-\bk') \rangle \nonumber \\
&=\frac{1}{N}\sum_{\bk'}\sum_{34} v_{14,32}(\bk-\bk') F_{43}(\bk'). \label{eq:a11}
\end{align}
Substituting Eq.\,\eqref{eq:vpp} to \eqref{eq:a11}, we obtain
\begin{subequations}
\begin{align}
& \hat{\Delta}(\bk)=\sum_\varGamma \sum_i \varDelta^\varGamma_i \hat{\varphi}^\varGamma_i(\bk), \\
& \varDelta^\varGamma_i=
v^\varGamma\frac{1}{N}\sum_\bk\sum_{12} F_{12}(\bk)\big[\hat{\varphi}^{\varGamma}_i(\bk)\big]^*_{12}. \label{eq:gap1}
\end{align}
\end{subequations}
Just below the transition temperature $T=T_c$, we can linearize $F_{12}(\bk)$ as 
\begin{align}
F_{12}(\bk)=T\sum_n \big[\hat{G}(\bk,i\omega_n)\hat{\Delta}(\bk)\hat{G}^*(-\bk,i\omega_n)\big]_{12},\label{eq:fk}
\end{align}
with Matsubara frequency $\omega_n=\pi T (2n+1)$. 
The one-particle normal Green's function $\hat{G}(\bk,i\omega_n)$ meets a similar relation to Eq.\,\eqref{eq:uhu},
\begin{align}
\hat{U}(g;\bk)\hat{G}(\bk,i\omega_n)\hat{U}^\dag(g;\bk)=\hat{G}(p\bk,i\omega_n).
\end{align}
Finally, from Eqs.\,\eqref{eq:gap1}, \eqref{eq:fk}, and the grand orthogonal theorem, we obtain the gap equations as follows,
\begin{align}
\begin{split}
\varDelta^\varGamma_i 
&=v^\varGamma\varDelta_i^{\varGamma}\frac{1}{d_{\varGamma}}\frac{T}{N}\sum_\bk\sum_j \sum_n \\
&~~{\rm Tr}\big[\hat{G}(\bk,i\omega_n)\hat{\varphi}^{\varGamma}_j(\bk)\hat{G}^*(-\bk,i\omega_n)\hat{\varphi}^{\varGamma\dag}_j(\bk)\big].\label{eq:gap2}
\end{split}
\end{align}
It should be noted that the gap equation \eqref{eq:gap2} is decoupled in each $\varGamma$, and also does not depend on the label $i$. 
This fact means that the gap function just below $T_c$ can be classified according to IRs of P in both symmorphic and non-symmorphic systems. In practice, $\hat{\varphi}^\varGamma_i(\bk)$ may be a linear combination of several basis functions in the same IR, namely, $\hat{\varphi}^\varGamma_i(\bk)=\sum_\alpha C_{\varGamma\alpha} \hat{\varphi}^\varGamma_{\alpha,i}(\bk)$. After diagonalizing the matrix $v^\varGamma_{\alpha\beta} =  v^\varGamma C_{\varGamma\alpha} C_{\varGamma\beta}^*$, $H_{\rm int}$ takes the form of Eq.\,\eqref{eq:intchannel} in the main text. The generalization to such situations is straightforward.

\subsection{Classification in symmorphic systems}\label{app:1}
In a symmorphic space group, apart from the lattice translations T, all generating symmetry operations leave at least one common point fixed. The generators consist of the elements in the semi-direct product of T and the point group P~\cite{bradley1}. In this case, for all point group operations $p=\{p|0\}\in {\rm P}$, we can always set $\hat{U}(p;\bk)$ in Eq.\,\eqref{eq:csym} to be $\bk$-independent $\hat{U}(p)$. This can be verified by the following discussions.

Let us denote $c^\dagger_{\ell\alpha b}({\bm r})$ as the electron creation operator, where $\ell$ indicates a basis function labeled by an IR of $\rm P$, $\alpha$ and $b$ denote the Kramers degrees of freedom and the position of the atom within a unit cell, respectively. $\br$ represents the position for the unit cell (lattice vector) and we also define the relative position for the $b$-atom $\br_b$ in a unit cell. In general, space group operations exchange the equivalent atoms in the same or the different unit cells. Considering the Fourier transform,
\begin{align}
c_{\ell\alpha b}^\dagger(\bk)=\frac{1}{\sqrt{N}}\sum_\br c_{\ell\alpha b}^\dag(\br) \exp[i\bk\cdot(\br+\br_b)],\end{align}
we can see the symmetry property of $c_{\ell\alpha b}^\dagger(\bk)$;
\begin{align}
g\,c_{\ell\alpha b}^\dagger(\bk)\,g^{-1}=e^{-ip\bk\cdot{\bm a}}\sum_{\alpha'b'}c_{\ell \alpha' b'}^\dagger(p\bk)D'_{b'b}(p)D''_{\alpha'\alpha}(p),\label{eq:cdd}
\end{align}
where $g=\{p|{\bm a}\}\in$ G. Here, $D'(p)$ and $D''(p)$ are the unitary matrices corresponding to the exchange of equivalent atoms and the rotation of the Kramers degrees of freedom, respectively. Since the phase factor $e^{-ip\bk\cdot{\bm a}}$ in Eq.\,\eqref{eq:cdd} is irrelevant to the point group operations alone, for all $p\in {\rm P}$, $\hat{U}(p;\bk)$ appearing in Eq.\,\eqref{eq:csym} becomes $\bk$-independent.

Equation\,\eqref{eq:cdd} also indicates that $c_{\ell\alpha b}^\dagger(\bk)$ is a basis function for a reducible representation of $\rm P$ regarding $c_{\ell\alpha b}^\dagger(\bk)\overset{p}{\mapsto}p\,c_{\ell\alpha b}^\dagger(p^{-1}\bk)p^{-1}$ as the action of $p$. Therefore, in the usual manner, we can construct the basis functions of the IRs of P from $c_{\ell\alpha b}^\dagger(\bk)$, by using the projection method. The obtained basis $c^\dag_{\varGamma i}(\bk)$ satisfies,

\begin{align}
p\,c^\dag_{\varGamma i}(\bk)\,p^{-1}
=\sum_{j}c^\dag_{\varGamma j}(p\bk)\big[\hat{D}^{(\varGamma)}(p)\big]_{ji}, \label{eq:phase}
\end{align}
where $\varGamma$ and $i$ are the IR of P and its basis, respectively. $\hat{D}^{(\varGamma)}(p)$ is the corresponding representation matrix. Here, we omit the other labels for simplicity. Due to the unitarity of the irreducible decomposition, we can always rewrite the Hamiltonian in the new basis $c^\dag_{\varGamma i}(\bk)$.

By using $c^\dag_{\varGamma i}(\bk)$ given above, Eq.\,\eqref{eq:hami} can be divided into each block for IRs of P,
\begin{align}
H_0=\sum_\bk\sum_{\varGamma_1\varGamma_2}\sum_{ij}
\big[\hat{h}(\bk;\!\varGamma_1\varGamma_2)\big]_{ij}
c^\dag_{\varGamma_1 i}(\bk)c_{\varGamma_2 j}(\bk),
\end{align}
where $\hat{h}(\bk;\!\varGamma_1\varGamma_2)$ satisfies
\begin{align}
\hat{h}(p\bk;\!\varGamma_1\varGamma_2)=\hat{D}^{(\varGamma_1)}(p)
\hat{h}(\bk;\!\varGamma_1\varGamma_2)\hat{D}^{(\varGamma_2)\dag}(p). \label{eq:hsym}
\end{align}
Similarly, Eq.\,\eqref{eq:g1} leads to, 
\begin{align}
&\Psi^\varGamma_i=\sum_\bk\sum_{\varGamma_1\varGamma_2}\sum_{j_1j_2}
\big[\hat{\varphi}^\varGamma_i(\bk;\varGamma_1\varGamma_2)\big]_{j_1j_2}c^\dag_{\varGamma_1j_1}(\bk)c^\dag_{\varGamma_2j_2}(-\bk), \label{eq:h1sym} \\
&\hat{D}^{(\varGamma_1)}(p)\hat{\varphi}^\varGamma_i(p^{-1}\bk;\varGamma_1\varGamma_2)(\hat{D}^{(\varGamma_2)}(p))^T \nonumber \\
&\hspace{30mm}=\sum_j \hat{\varphi}^\varGamma_j(\bk;\varGamma_1\varGamma_2)\mathcal{D}_{ji}^{(\varGamma)}(p). \label{eq:psym}
\end{align}
Equation\,\eqref{eq:psym} indicates that 
$\hat{\varphi}^\varGamma_i(\bk;\varGamma_1\varGamma_2)$ with $\varGamma$ IR can be obtained from the subduction \mbox{$\varGamma_\bk\otimes(\varGamma_1\!\otimes\!\varGamma_2)\!\downarrow\!{\rm P}$} [See Eq.\;\eqref{eq:subduc}], where $\varGamma_\bk$ denotes the IR of the momentum transform: $\hat{\varphi}^\varGamma_i(\bk)\overset{p}{\mapsto}\hat{\varphi}^\varGamma_i(p^{-1}\bk)$.

Note that Eq.\,\eqref{eq:hsym} is similar to the case of $\varGamma=A_{1g}$ in Eq.\,\eqref{eq:psym}, apart from the IR for the Kramers sector. It is given by $\varGamma_1\otimes\varGamma_2$ for \eqref{eq:psym}, while $\varGamma_1\otimes\varGamma_2^*$ for \eqref{eq:hsym}.
Therefore, the tables shown in the present paper will be helpful also in constructing a generic tight-binding model in multi-orbital systems.

Finally, let us comment on non-symmorphic systems. In this case, the above discussion is no longer applicable due to inevitable $\bk$ dependence in the phase factor of $\hat{U}(g;\bk)$. An available alternative method~\cite{izyumov1,yarzhemsky1,yarzhemsky2,micklitz1} is the classification based on a little group at a given $\bk$ point. 
This is applicable in both symmorphic and non-symmorphic systems, but beyond the scope of the present paper and we leave it as a future study.

\section{Basis functions in double-valued representations}\label{app:2}
In this Appendix, we list some basis functions for double-valued IRs in O, D$_4$ and D$_6$ group. 
In the list below, $\ket{j;j_z}$ represents the basis of the total angular momentum $j$ and the $z$ component $j_z$ in SU(2) symmetry group.
\begin{itemize}
\item O group
\begin{align}
& \Ket{\varGamma_7;\pm}=\sqrt{\frac{1}{6}}\Ket{\frac{5}{2};\pm\frac{5}{2}}-\sqrt{\frac{5}{6}}\Ket{\frac{5}{2};\mp\frac{3}{2}}, \nonumber \\
& \Ket{\varGamma_{8a};\pm}=\sqrt{\frac{5}{6}}\Ket{\frac{5}{2};\pm\frac{5}{2}}+\sqrt{\frac{1}{6}}\Ket{\frac{5}{2};\mp\frac{3}{2}}, \nonumber \\
& \Ket{\varGamma_{8b};\pm}=\Ket{\frac{5}{2};\pm\frac{1}{2}}, \\[5mm]
& \Ket{\varGamma_{8a};\pm}=\pm\Ket{\frac{3}{2};\mp\frac{3}{2}}, \nonumber \\
& \Ket{\varGamma_{8b};\pm}=\pm\Ket{\frac{3}{2};\pm\frac{1}{2}}.
\end{align}
\item D$_4$ group
\begin{align}
&\Ket{\varGamma_6;\pm}=\Ket{\frac{5}{2};\pm\frac{1}{2}}, \nonumber \\
&\Ket{\varGamma_7;\pm}=\cos\theta\Ket{\frac{5}{2};\pm\frac{5}{2}}+\sin\theta\Ket{\frac{5}{2};\mp\frac{3}{2}}, \\[5mm]
&\Ket{\varGamma_6;\pm}=\mp\Ket{\frac{3}{2};\pm\frac{1}{2}}, \nonumber \\
&\Ket{\varGamma_7;\pm}=\mp\Ket{\frac{3}{2};\mp\frac{3}{2}}. 
\end{align}
\item D$_6$ group
\begin{align}
&\Ket{\varGamma_7;\pm}=\Ket{\frac{5}{2};\pm\frac{1}{2}},~
&\Ket{\varGamma_8;\pm}=\Ket{\frac{5}{2};\pm\frac{5}{2}}, \nonumber \\
&\Ket{\varGamma_9;\pm}=\Ket{\frac{5}{2};\mp\frac{3}{2}}, \\[5mm]
&\Ket{\varGamma_7;\pm}=\mp\Ket{\frac{3}{2};\pm\frac{1}{2}}, \nonumber \\
&\Ket{\varGamma_9;\pm}=\mp\Ket{\frac{3}{2};\mp\frac{3}{2}}.
\end{align}
\end{itemize}
In SI invariant systems, these basis functions are classified into even or odd parity, following the orbital angular momentum $\ell=j\mp s$ with the spin $s=1/2$. 
Under the time-reversal operation $\varTheta$, we take the following convention,
\begin{align}
\varTheta\Ket{j;j_z}=(-1)^{j+j_z}\Ket{j;-j_z},
\end{align}
and thus, the basis functions defined above meet 
$\varTheta\ket{\varGamma;\pm}=\mp\ket{\varGamma;\mp}$ for any $\varGamma$. 

\section{Symmetry argument of band-based representation}
Here, we describe a procedure to fix the U(2) phase ambiguity in the band-based representation, and demonstrate that the gap structure looks apparently different, depending on the choice of the fixed phase, although the structure of excitations is unchanged.

\subsection{Phase fixing procedure}\label{app:3}
Let us consider an $N$-orbital system. If all the orbitals are independent and not hybridized with each other, then any electron in the band representation consists of single orbital; a unitary matrix $u(\bk)$ is an identity matrix. No matter how complicated the band structure is, we can line up orbital indices in such a way that the dominant orbital in each band is arranged in a diagonal position of the matrix $u(\bk)$.
After this procedure, we now fix the U(2) gauge. 

Under the presence of the SI and TR symmetries, the following relation holds
\begin{align}
(\varTheta I)\,c^\dag_{n\sigma}(\bk)\,(\varTheta I)^{-1}=\sum_{\sigma'}c^\dag_{n\sigma'}(\bk)(i\sigma^y)_{\sigma'\sigma}. \label{eq:a3-1}
\end{align}
Substituting Eq.\,\eqref{eq:umat} into the both sides of Eq.\,\eqref{eq:a3-1}, we obtain,
\begin{align}
u_{\ell\alpha,n\sigma}(\bk) = (-1)^{P_\ell}\sum_{\alpha'\sigma'}(i\sigma^y)_{\alpha\alpha'}u^*_{\ell\alpha',n\sigma'}(\bk)(i\sigma^y)_{\sigma'\sigma}^\dag, \label{eq:a3-2}
\end{align}
where $P_\ell$ is the parity of the orbital $\ell$. In what follows, we focus on the $2\times 2$ submatrix $\hat{u}(\bk;\ell n)$, where $\big[\hat{u}(\bk;\ell n)\big]_{\alpha\sigma}\equiv u_{\ell\alpha,n\sigma}(\bk)$. From Eq.\,\eqref{eq:a3-2}, we find that each submatrix $\hat{u}(\bk;\ell n)$ satisfies,
\begin{align}
\hat{u}(\bk;\ell n)\hat{u}^\dagger(\bk;\ell n)
=|\det \hat{u}(\bk;\ell n)| I_{2\times 2}, \label{eq:ukram}
\end{align}
which is independent of $P_\ell$. Here, $I_{2\times2}$ is the $2\times2$ identity matrix. 
Let us consider the following matrix,
\begin{align}
\hat{K}_n(\bk)=\frac{1}{\sqrt{|\det \hat{u}(\bk;n n)|}}\hat{u}(\bk;n n).
\end{align}
Then, the U(2) phase ambiguity can be fixed by redefining the unitary matrix as follows,
\begin{align}
\tilde{u}_{\ell\alpha,n\sigma}(\bk)=\big[\hat{u}(\bk;\ell n)\hat{K}_n^\dag(\bk)\big]_{\alpha\sigma}.
\end{align}
Indeed, this matrix diagonalizes $H_0$, and the phase for $\ell=n$ component is fixed to be positive real as,
\begin{align}
\tilde{u}_{n\alpha,n\sigma}(\bk)=\sqrt{|\det \hat{u}(\bk;n n)|}\delta_{\alpha\sigma}. 
\end{align}
In the main text, $u(\bk)$ means this $\tilde{u}(\bk)$, unless otherwise noted.
It should be noted that the unitary matrix preserves the Kramers label $\alpha$ for each orbital. In other words, if the $n$th orbital belongs to a $\varGamma$ IR, the corresponding band electron also belongs to the same $\varGamma$ IR. It is useful to discuss the nodal positions in the band-based gap functions as will be shown in Appendix \ref{app:4} and \ref{app:5}. In addition, the unitary matrix obtained in the above way smoothly connects to the $2N\times2N$ identity matrix in the limit where there is no hybridization between different orbitals, which is one of desirable properties as a diagonalizing matrix.

Note that the gap structure in the multi-orbital systems strongly depends on the way of the phase fixing, although observable quantities are unchanged. Depending on the way, meaningless complicated structure can appear in the obtained gap structure. We will demonstrate this point in Appendix \ref{app:5}.

\subsection{Symmetry of unitary matrix and band-based pair amplitude}\label{app:4}
Here, let us study the symmetry of the unitary matrix $u_{\ell\alpha,n\sigma}(\bk)$. 
In our case, due to the phase fixing mentioned in Appendix \ref{app:3}, we can explicitly discuss the symmetry. 
In actual calculations, we first diagonalize $H_0$ in the irreducible Brillouin zone (BZ). At this stage, the obtained unitary matrix still has an arbitrary phase. Then, we fix the phase, following the procedure explained in Appendix \ref{app:3}. The unitary matrix in the whole first BZ can be obtained by the following transformation,
\begin{align}
\hat{u}(p\bk;\ell n)=\hat{D}^{(\varGamma_\ell)}(p)\hat{u}(\bk;\ell n)\hat{D}^{(\varGamma_n)\dag}(p),\label{eq:usdef}
\end{align}
where $\varGamma_\ell (\varGamma_n)$ denotes IRs of $\ell (n)$, and $\bk$ is in the irreducible BZ.
Note that Eq.\,\eqref{eq:usdef} is similar to Eq.\,\eqref{eq:hsym}. This indicates that our unitary matrix has the same structure as $\hat{h}(\bk)$ with respect to the symmetry. 
From this property, the symmetry of the band-based gap functions is readily available from that of orbital-based ones.

Indeed, using Eqs.\,\eqref{eq:phase} and \eqref{eq:usdef}, we obtain
\begin{align}
p\,\tilde{c}^\dag_{n\sigma}(\bk)\,p^{-1}
&=\sum_{\ell\alpha\alpha'}c_{\ell\alpha}^\dag(p\bk)
\big[\hat{D}^{(\varGamma_\ell)}(p)\big]_{\alpha\alpha'} u_{\ell\alpha',n\sigma}(\bk) \nonumber\\
&=\sum_{\ell\alpha\sigma'}c_{\ell\alpha}^\dag(p\bk)
u_{\ell\alpha,n\sigma'}(p\bk)\big[\hat{D}^{(\varGamma_n)}(p)\big]_{\sigma'\sigma}\nonumber\\
&=\sum_{\sigma'}\tilde{c}^\dag_{n\sigma'}(p\bk)\big[\hat{D}^{(\varGamma_n)}(p)\big]_{\sigma'\sigma}.
\end{align}
This transformation property for the band $n$ is the same as the orbital-based case in Eq.\,\eqref{eq:phase}.
Therefore, when we consider the band-based pair amplitude,
\begin{align}
\tilde{F}_{n\sigma,n'\sigma'}(\bk)\equiv \langle \tilde{c}_{n\sigma}(\bk) \tilde{c}_{n'\sigma'}(-\bk)
\rangle, 
\end{align}
the symmetry arguments in Sec.\,\ref{sec:2} hold for this band-based gap functions. Also it is evident that Tables \ref{table2-O}-\ref{table2-D6} are valid.
However, as mentioned in the main text, such band-based arguments are insufficient to understand a variety of multi-orbital superconductivity, because the pairing interactions can be more clearly defined in the orbital-based representation. 
Indeed, in the band-based representation, we will miss the presence of additional nodes as discussed in Sec.\,\ref{sec:3}, which are not symmetry-protected but inevitable from the orbital-based viewpoint. Thus, it is clear that the unitary matrix $u_{\ell\alpha,n\sigma}(\bk)$ can possess significant information about $\bk$ dependence of gap functions.

\subsection{Efficacy of the phase fixing}\label{app:5}
\begin{figure}[b]
\centering
\includegraphics[width=8.5cm]{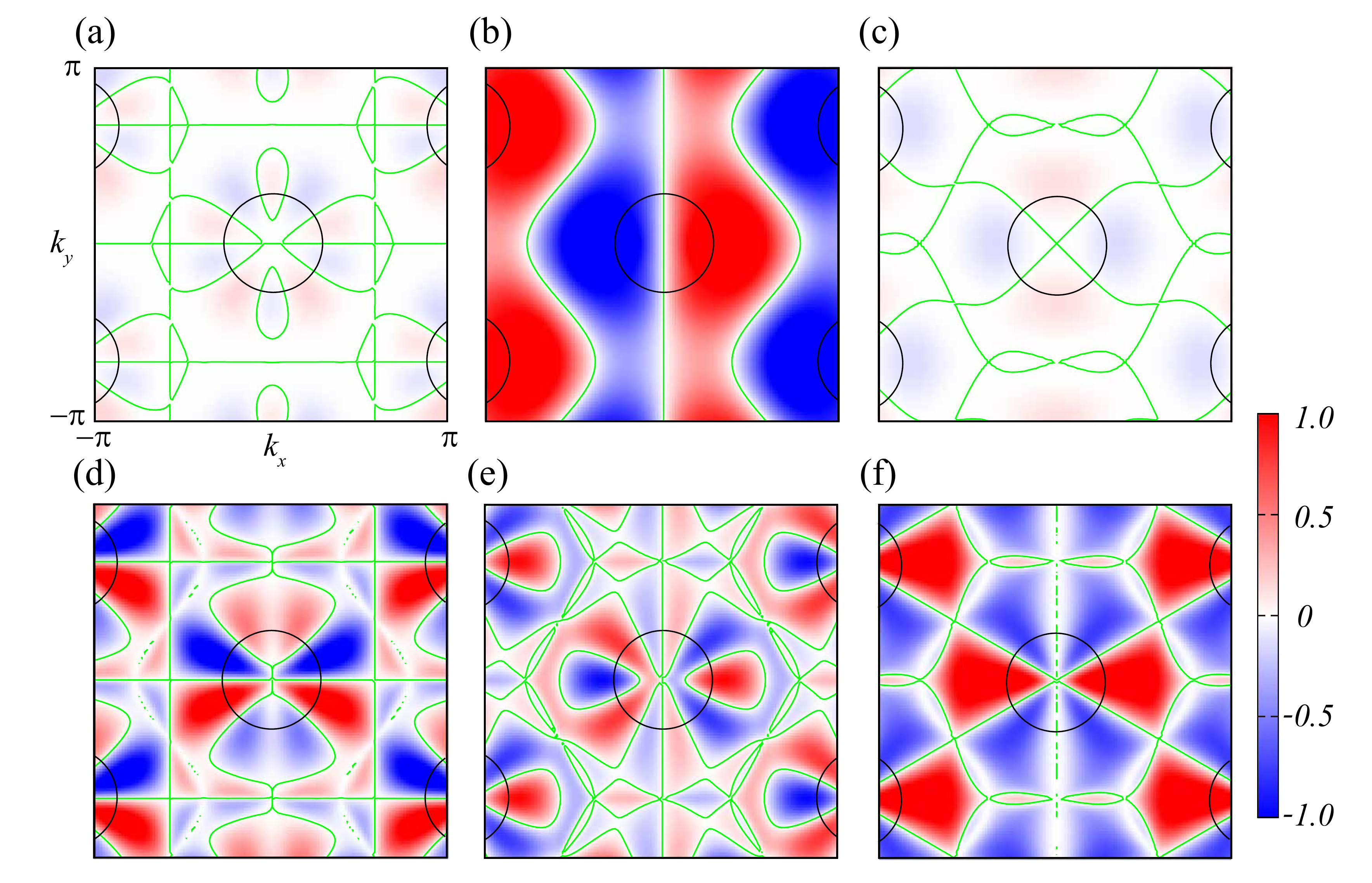}
\caption{(Color online) Band-based gap functions of the lower band in $k_z=\pi$ plane. (a) $d_x$, (b) $d_y$, and (c) $d_z$ components of the band-based gap functions are obtained by our phase fixing procedure, where the Kramers index is labeled by that of the major $\varGamma_{9g}$ orbital. (d) $d_x$, (e) $d_y$, and (f) $d_z$ components of the gap functions, labeled by the Kramers index of the minor $\varGamma_{7g}$ orbital. Green dashed lines denote gap nodes. }
\label{fig:app}
\end{figure}

Finally, let us demonstrate an advantage of our phase fixing method. 
We consider a two-orbital model constructed from $\varGamma_{7g}$ and $\varGamma_{9g}$ orbitals in $\rm D_{6h}$ group.
The general form of $\hat{h}(\bk)$ in Eq.\,\eqref{eq:hami} is given as
\begin{align}
\begin{split}
\hat{h}(\bk)&=h^{A_{1g}}_1\tau^0\sigma^0+h^{A_{1g}}_2\tau^z\sigma^0+h^{E_{1g}}_1\tau^y\sigma^x \\
& -h^{E_{1g}}_2\tau^x\sigma^y+h^{E_{2g}}_1\tau^y\sigma^z+h^{E_{2g}}_2\tau^x\sigma^0,
\end{split}
\end{align}
where $h^\varGamma_{1,2}$ consists of basis functions of $\varGamma$ IRs:
\begin{align*}
& h^{A_{1g}}_1=-t_0\Big(\!\cos \sqrt{3}k_x \!+\! 2\cos \frac{\sqrt{3}k_x}{2} 
\cos \frac{3k_y}{2}\!\Big)-\mu, \\
& h^{A_{1g}}_2=-t_1, ~~ h^{E_{1g}}_1=t_2 s_x' \sin k_z, ~~ h^{E_{1g}}_2=t_2 s_y'\sin k_z, \\
& h^{E_{2g}}_1=2t_3 s_x's_y', ~~~ h^{E_{1g}}_2=t_3 (s_x'^2-s_y'^2),
\end{align*}
with
\begin{align*}
s_x' &=\sin \sqrt{3}k_x + \sin \frac{\sqrt{3}k_x}{2} \cos \frac{3k_y}{2}, \\
s_y' &= \sqrt{3}\sin \frac{3k_y}{2} \cos \frac{\sqrt{3}k_x}{2}.
\end{align*}
Here, we set $t_0$ to the unit of energy and $(t_1,t_2,t_3)=(0.25,0.05,0.05)$ and $\mu=-1.20$. With these parameters, the dominant component of the lower (upper) band is almost composed of $\varGamma_{7g} (\varGamma_{9g})$ orbital. Below, we will focus on the band mainly composed of $\varGamma_{9g}$ and will not discuss the other band for simplicity.

For example, let us consider one of $E_{2u}$ pairing states in $\varGamma_{9g}$ orbital, i.e., $\phi^{E_{1u}}_1\by$ for $\varGamma_9\otimes\varGamma_9$ pairs in Table \ref{table2-D6}: 
\begin{align}\label{eq:orbgap}
\hat{\varphi}^{E_{2u}}(\bk)=\phi^{E_{1u}}_1(\bk)(\tau^0-\tau^z){\bm y},
\end{align}
with $\phi^{E_{1u}}_1(\bk)=s'_x$. 
Here, $\tau^0-\tau^z$ represents the pair in $\varGamma_9\otimes\varGamma_9$. In Fig.\,\ref{fig:app}, we illustrate the band-based gap function for the lower band, which is evaluated via Eq.\,\eqref{eq:uuF}. Figures \ref{fig:app}(a), (b) and (c) depict, respectively, $d_x$, $d_y$, and $d_z$ components with our phase fixing method, where the upper (lower) band is smoothly connected with the $\varGamma_{7g}\, (\varGamma_{9g})$ orbital. Through the unitary matrix, $d_x$ and $d_z$ components are induced, but the magnitude is very small. $d_y$ component is almost the same as $\phi^{E_{1u}}_1(\bk)$ given in Eq.\,\eqref{eq:orbgap}. In contrast, one can see the complicated gap structures in Figs.\,\ref{fig:app}(d)-(f), the magnitudes of which are comparable to each other. Here, the Kramers index for the lower band is labeled by that for the minor $\varGamma_{7g}$ orbital. At a glance, there seem to exist complicated additional nodes. The gap amplitude $\sqrt{|{\bm d}|}$, however, is identical to that shown in Figs.\,\ref{fig:app}(a)-(c), and is independent of the way of the phase fixing.
This demonstrates that our phase fixing method is effective and useful in the discussion about the gap structures in the multi-orbital systems.

\end{appendix}


\begin{thebibliography}{99}
\bibitem{BCS} J. Bardeen, L. N. Cooper, and J. R. Schrieffer, Phys. Rev. {\bf 108}, 1175 (1957).
\bibitem{Stewart} G. R. Stewart, Rev. Mod. Phys. {\bf 56}, 755 (1984).
\bibitem{cuprate} W. E. Pickett, Rev. Mod. Phys. {\bf 61}, 433 (1989).
\bibitem{Miyake} K. Miyake, S. Schmitt-Rink, and C. M. Varma, Phys. Rev. B {\bf 34}, 6554 (1986).
\bibitem{Scalapino} D. J. Scalapino, E. Loh, Jr., and J. E. Hirsch, Phys. Rev. B, {\bf 34}, 8190 (1986).
\bibitem{Legget} A. J. Leggett, Rev. Mod. Phys. {\bf 47}, 331 (1975). 
\bibitem{ce115A} K. Izawa, H. Yamaguchi, Y. Matsuda, H. Shishido, R. Settai, and Y. Onuki, Phys. Rev. Lett. {\bf 87}, 057002 (2001).
\bibitem{ce115B} K. An, T. Sakakibara, R. Settai, Y. Onuki, M. Hiragi, M. Ichioka, and K. Machida, Phys. Rev. Lett. {\bf 104}, 037002 (2010).
\bibitem{Harlingen} D. J. Van Harlingen, Rev. Mod. Phys. {\bf 67}, 515 (1995).
\bibitem{volovik1} G. E. Volovik and L. P. Gor'kov, JETP Lett. {\bf 39}, 674 (1984).
\bibitem{volovik2} G. E. Volovik and L. P. Gor'kov, Sov. Phys. JETP {\bf 61}, 843 (1985).
\bibitem{ueda1} K. Ueda and T. M. Rice, Phys. Rev. B {\bf 31}, 7114 (1985).
\bibitem{sigrist1} M. Sigrist and K. Ueda, Rev. Mod. Phys. {\bf 63}, 239 (1991).
\bibitem{salus1} J. A. Sauls, Adv. Phys. {\bf 43}, 113 (1994).
\bibitem{joynt1} R. Joynt and L. Taillefer, Rev. Mod. Phys. {\bf 74}, 235 (2002).
\bibitem{annet1} J. F. Annett, Adv. Phys. {\bf 39}, 83 (1990).
\bibitem{rice1} T. M. Rice and M. Sigrist, J. Phys.: Condens. Matter {\bf 7}, L643 (1995).
\bibitem{machida1} K. Machida, M. Ozaki, and T. Ohmi, J. Phys. Soc. Jpn. {\bf 65}, 3720 (1996).
\bibitem{mackenzie1} A. P. Mackenzie and Y. Maeno, Rev. Mod. Phys. {\bf 75}, 657 (2003).
\bibitem{bauer1} E. Bauer, G. Hilscher, H. Michor, Ch. Paul, E. W. Scheidt, A.
Gribanov, Yu. Seropegin, H. No$\ddot{\rm e}$l, M. Sigrist, and P. Rogl, Phys. Rev. Lett. {\bf 92}, 027003 (2004).
\bibitem{akazawa1} T. Akazawa, H. Hidaka, T. Fujiwara, T. C. Kobayashi, E. Yamamoto, Y. Haga, R. Settai, and Y. Onuki, J. Phys.: Condens. Matter {\bf 16}, L29 (2004).
\bibitem{Goll} G. Goll, M. Marz, A. Hamann, T. Tomanic, K. Grube, T. Yoshino, and T. Takabatake, Physica B {\bf 403}, 1065 (2008).
\bibitem{Edelstein} V. M. Edelstein, Sov. Phys. JETP {\bf 68}, 1244 (1989). 
\bibitem{gorkov1} L. P. Gor'kov and E. I. Rashba, Phys. Rev. Lett. {\bf 87}, 037004 (2001).
\bibitem{Yip0} S. K. Yip, Phys. Rev. B {\bf 65}, 144508 (2002).
\bibitem{frigeri1} P. A. Frigeri, D. F. Agterberg, A. Koga, and M. Sigrist, Phys. Rev. Lett. {\bf 92}, 097001 (2004).
\bibitem{sergienko1} I. A. Sergienko and S. H. Curnoe, Phys. Rev. B {\bf 70}, 214510 (2004).
\bibitem{sato} M. Sato and S. Fujimoto, Phys. Rev. B {\bf 79}, 094504 (2009).
\bibitem{upt3rev1} J. A. Sauls, Adv. Phys. {\bf 43}, 113 (1994).
\bibitem{upt3rev2} R. Joynt and L. Taillefer, Rev. Mod. Phys. {\bf 74}, 235 (2002).
\bibitem{micklitz1} T. Micklitz and M. R. Norman, Phys. Rev. B {\bf 80}, 100506(R) (2009).
\bibitem{UPt3} T. Nomoto and H. Ikeda, unpublished.
\bibitem{izyumov1} Y. A. Izyumov, V. M. Laptev, and V. N. Syromyatnikov, Int. J. Mod. Phys. B {\bf 3}, 1377 (1989).
\bibitem{yarzhemsky1} V. G. Yarzhemsky and E. N. Murav'ev, J. Phys.: Condens. Matter {\bf 4}, 3525 (1992).
\bibitem{yarzhemsky2} V. G. Yarzhemsky, Phys. Status Solidi B {\bf 209}, 101 (1998).
\bibitem{bradley1} C. J. Bradley and A. P. Cracknell, in {\it The Mathematical Theory of Symmetry in Solids} (Oxford University Press, Oxford, 1972).
\bibitem{lee1} P. A. Lee and X.-G. Wen, Phys. Rev. B {\bf 78}, 144517 (2008).
\bibitem{hu1} J. Hu and N. Hao, Phys. Rev. X {\bf 2}, 021009 (2012).
\bibitem{ong1} T. T. Ong and P. Coleman, Phys. Rev. Lett. {\bf 111}, 217003 (2013).
\bibitem{zhou1} Y. Zhou, W.-Q. Chen, and F.-C. Zhang, Phys. Rev. B {\bf 78}, 064514 (2008).
\bibitem{wan1} Y. Wan and Q.-H. Wang, Europhys. Lett. {\bf 85} 57007 (2008).
\bibitem{you1} W.-L. You, S.-J. Gu, G.-S. Tian, and H.-Q. Lin, Phys. Rev. B {\bf 79}, 014508 (2009).
\bibitem{daghofer1} M. Daghofer, A. Nicholson, A. Moreo, and E. Dagotto, Phys. Rev. B {\bf 81}, 014511 (2010).
\bibitem{fischer1} M. H. Fischer, New J. Phys. {\bf 15}, 073006 (2013).
\bibitem{shiina1} R. Shiina, H. Shiba, and P. Thalmeier, J. Phys. Soc. Jpn. {\bf 66}, 1741 (1997).
\bibitem{Yip} S. Yip and A. Garg Phys. Rev. B {\bf 48}, 3304 (1993).
\bibitem{memo0} We have used the word ``spin singlet/triplet'' according to the usual convention, although this means antisymmetrized/symmetrized representations for the pairs of Kramers degrees of freedom.
\bibitem{HKim} H. Kim, K. Wang, Y. Nakajima, R. Hu, S. Ziemak, P. Syers, L. Wang, H. Hodovanets, J. D. Denlinger, P. M. R. Brydon, D. F. Agterberg, M. A. Tanatar, R. Prozorov, and J. Paglione, arXiv:1603.03375.
\bibitem{Brydon} P. M. R. Brydon, L. Wang, M. Weinert, and D. F. Agterberg, Phys. Rev. Lett. {\bf 116}, 177001 (2016).
\bibitem{memo1} In our classification, the $p$-wave septet pairing state proposed in Refs.~\cite{HKim,Brydon} corresponds to one of the $A_2$ pairings in $\varGamma_8\otimes\varGamma_8$: $k_1\eta^1\bx+k_2\eta^2\by+k_3\eta^3\bz$ in Table \ref{table2-O}.
\bibitem{Bishop} C. B. Bishop, G. Liu, E. Dagotto, and A. Moreo, arXiv:1602.02420.
\bibitem{Kontani2} H. Kontani, Phys. Rev. B {\bf 70}, 054507 (2004).
\bibitem{Matsubayashi} K. Matsubayashi, T. Tanaka, A. Sakai, S. Nakatsuji, Y. Kubo, and Y. Uwatoko, Phys. Rev. Lett. {\bf 109}, 187004 (2012).
\bibitem{Tsujimoto} M. Tsujimoto, Y. Matsumoto, T. Tomita, A. Sakai, and S. Nakatsuji, Phys. Rev. Lett. {\bf 113}, 267001 (2014).
\bibitem{Effantin} J. M. Effantin, J. Rossat-Mignod, P. Burlet, H. Bartholin, S. Kunii, and T. Kasuya, J. Magn. Magn. Mater. {\bf 47-48}, 145 (1985).
\bibitem{Tayama} T. Tayama, T. Sakakibara, K. Kitami, M. Yokoyama, K. Tenya,
H. Amitsuka, D. Aoki, Y. \=Onuki, and Z. Kletowski, J. Phys. Soc. Jpn. {\bf 70} 248 (2001).
\bibitem{OnimaruPrPb3} T. Onimaru, T. Sakakibara, N. Aso, H. Yoshizawa, H. S. Suzuki, and T. Takeuchi, Phys. Rev. Lett. {\bf 94}, 197201 (2005).
\bibitem{Onimaru} T. Onimaru, K. T. Matsumoto, Y. F. Inoue, K. Umeo, T. Sakakibara, Y. Karaki, M. Kubota, and T. Takabatake, Phys. Rev. Lett. {\bf 106}, 177001 (2011).
\bibitem{MatsuUnpub} K. Matsubayashi {\it et al.}, unpublished.
\bibitem{quadrupole} P. Morin and D. Schmitt, in {\it Ferromagnetic Materials}, edited by K. H. J. Buschow and E. P. Wohlfarth (Elsevier, Amsterdam, 1990) Vol. 5, p. 1.
\bibitem{mizuguchi1} Y. Mizuguchi, J. Phys. Chem. Solids {\bf 84}, 34 (2015).
\bibitem{usui1} H. Usui, K. Suzuki, and K. Kuroki, Phys. Rev. B {\bf 86}, 220501(R) (2012).
\bibitem{lee2} J. Lee, M. B. Stone, A. Huq, T. Yildirim, G. Ehlers, Y. Mizuguchi, O. Miura, Y. Takano, K. Deguchi, S. Demura, and S.-H. Lee, Phys. Rev. B {\bf 87}, 205134 (2013).
\bibitem{athauda1} A. Athauda, J. Yang, S. Lee, Y. Mizuguchi, K. Deguchi, Y. Takano, O. Miura, and D. Louca, Phys. Rev. B {\bf 91}, 144112 (2015).
\bibitem{Kittaka} S. Kittaka, T. Sakakibara, M. Hedo, Y. Onuki, and K. Machida, J. Phys. Soc. Jpn. {\bf 82}, 123706 (2013).
\bibitem{Izawa} K. Izawa, Y. Nakajima, J. Goryo, Y. Matsuda, S. Osaki, H. Sugawara, H. Sato, P. Thalmeier, and K. Maki, Phys. Rev. Lett. {\bf 90}, 117001 (2003).
\bibitem{Matsuda} Y. Matsuda, K. Izawa, and I. Vekhter, J. Phys.: Condens. Matter {\bf 18}, R705 (2006).
\end{thebibliography}
\end{document}